\documentclass[a4paper,12pt]{article}
\usepackage{authblk}

\usepackage{framed,multirow}
\usepackage{amssymb}
\usepackage{caption}
\usepackage{textcomp}
\usepackage{latexsym}
\usepackage[explicit,noindentafter]{titlesec}
\usepackage{url}
\usepackage{xcolor}
\usepackage{graphicx}
\usepackage[T1]{fontenc}
\usepackage[utf8]{inputenc}
\usepackage{amsmath}
\usepackage{bm}
\usepackage{xfrac}
\usepackage[htt]{hyphenat}
\usepackage{empheq}
\usepackage{csquotes}
\usepackage{fancybox}
\usepackage[hidelinks]{hyperref}
\usepackage{fancyhdr}
\usepackage{xcolor}
\usepackage{soul}
\usepackage{soulutf8}
\usepackage{braket}
\usepackage{tikz}
\usepackage{pgf}
\usepackage{pgfplots}
\usepgfplotslibrary{groupplots}
\pgfplotsset{compat=newest}
\usepackage{listings}
\usepackage{pbox}
\usepackage{pdfcomment}

\definecolor{newcolor}{rgb}{.8,.349,.1}

\definecolor{colorb}{RGB}{0,0,0}
\definecolor{colorw}{RGB}{255,255,255}
\definecolor{color1}{RGB}{32,214,120}
\definecolor{color2}{rgb}{0.95,0.5,0.1}
\definecolor{color3}{rgb}{0.05,0.05,0.9}
\def\cterm1{\color{color1}}
\def\cterm2{\color{color2}}
\def\cterm3{\color{color3}}

\definecolor{colorrewier1}{RGB}{66, 134, 244}
\definecolor{colorrewier2}{RGB}{244, 110, 66}
\definecolor{colorrewier3}{RGB}{55, 166, 55}
\definecolor{colorrewierme}{RGB}{71, 163, 14}
\definecolor{coloreditor}{RGB}{21, 76, 200}
\definecolor{colorme}{RGB}{186, 52, 235}

\def\cblack{\color{black}} 
\def\ced{\color{black}}

\def\crevi{\color{black}} 
\def\crevii{\color{black}}
\def\creviii{\color{black}}
\def\crevme{\color{black}}

\def\wtextbullet{{\color{colorw}\textbullet}}

\definecolor{lightblue}{RGB}{219, 235, 252}
\definecolor{lightgreen}{RGB}{219, 252, 232}
\definecolor{lightorange}{RGB}{252, 237, 219}
\definecolor{lightpurp}{RGB}{252, 219, 219}
\definecolor{lightred}{RGB}{240, 219, 252}
\definecolor{lightyellow}{RGB}{248, 252, 219}

\newcommand{\stexttt}[1]{\small{\texttt{#1}}}
\newcommand{\spath}[1]{{\small\url{#1}}}

\def\bE{\ensuremath{\mathbf{E}}}
\def\sh{\ensuremath{h}}

\def\bn{\ensuremath{\mathbf{n}}}
\def\br{\ensuremath{\mathbf{r}}}
\def\bx{\ensuremath{\mathbf{x}}}
\def\by{\ensuremath{\mathbf{y}}}
\def\bz{\ensuremath{\mathbf{z}}}

\def\exp{\ensuremath{\mathrm{exp}}}

\def\epsinf{\ensuremath{\varepsilon_\infty}}
\def\omd{\ensuremath{\omega_d}}
\def\omfb{\ensuremath{\omega^{\diamond}}}
\def\gamd{\ensuremath{\gamma_d}}
\def\om{\ensuremath{\omega}}

\def\M{\boldsymbol{M}}
\def\U{\mathcal{U}}

\def\epz{\ensuremath{\epsilon_0}}
\def\muz{\ensuremath{\mu_0}}

\def\x{\boldsymbol{x}}
\def\r{\boldsymbol{r}}

\def\dfrac{\displaystyle\frac}
\def\dint{\displaystyle\int}
\def\curl{\ensuremath{\mathbf{curl}}}
\def\div{\mathrm{div}}
\def\grad{\ensuremath{\mathbf{grad}}}
\def\tensmur{\boldsymbol{\mu}_r}
\def\tensepsr{\boldsymbol{\varepsilon}_r}

\def\tensid{\boldsymbol{I}}
\def\tenspml{\boldsymbol{S}}

\def\i{i}
\def\AD{\boldsymbol{A}_1^d}
\newcommand{\eps}[2]{\varepsilon_{{#1},{#2}}}

\def\bE{\ensuremath{\mathbf{E}}}
\def\bH{\ensuremath{\mathbf{H}}}

\def\ofom{(\omega)}

\def\dfrac{\displaystyle\frac}
\def\dint{\displaystyle\int}

\def\bE{\ensuremath{\mathbf{E}}}
\def\bW{\ensuremath{\mathbf{W}}}
\def\bH{\ensuremath{\mathbf{H}}}

\def\M{\boldsymbol{M}}
\def\ofom{(\omega)}

\def\N{\mathcal{N}}
\def\D{\mathcal{D}}

\def\om{\omega}

\def\toslash{{\scriptscriptstyle\oslash}}

\newcommand{\Galerkin}[3]{\ensuremath{\int_{#3}#1\cdot\overline{#2}\,\mathrm{d}{\Omega}}}
\newcommand{\Galerkinl}[3]{\ensuremath{\int_{#3}#1\cdot\overline{#2}\,\mathrm{d}{\Gamma}}}

\def\flagdraftfig{false}
\newcommand{\Fig}[1]{Fig.~\ref{#1}}
\newcommand{\tabmr}[1]{\pbox{20cm}{\vspace{.2\baselineskip}#1\vspace{.2\baselineskip}}}

\title{{\ced Non-linear eigenvalue problems with GetDP and SLEPc:} Eigenmode computations of frequency-dispersive photonic open structures.}

\author[1,*]{Guillaume Dem{\'e}sy}
\author[1]{André Nicolet}
\author[1]{Boris Gralak}
\author[2]{Christophe Geuzaine}
\author[3]{Carmen Campos}
\author[3]{Jose E. Roman}

\begin{footnotesize}
\affil[1]{Aix Marseille Univ, CNRS, Centrale Marseille, Institut Fresnel, Marseille, France.}
\affil[2]{University of Liège, Dept. of Electrical Engineering and Computer Science, Montefiore Institute B28, Quartier Polytech 1, Allée de la Découverte 10, B-4000 Liège, Belgium.}
\affil[3]{Universitat Politècnica de València, D. Sistemes Informàtics i Computació, Camí de Vera, s/n, E-46022 València, Spain.}
\affil[*]{Corresponding author : \texttt{guillaume.demesy@fresnel.fr}.}
\end{footnotesize}

\begin{document}
\maketitle







\begin{abstract} 
    
{\ced We present a framework to solve non-linear eigenvalue problems suitable {\crevi for} a Finite Element discretization. The
implementation is based on the open-source finite element software GetDP and the open-source library SLEPc. As template
examples,} we propose and compare in detail different ways to address the numerical computation of the electromagnetic modes of
frequency-dispersive objects. This is a non-linear eigenvalue problem involving a non-Hermitian operator. A classical
finite element formulation is derived for five different solutions and solved using algorithms adapted to the large size
of the resulting discrete problem. The proposed solutions are applied to the computation of the dispersion relation of a
diffraction grating made of a Drude material. The important numerical consequences {\crevme linked to} the presence of sharp
corners and sign-changing coefficients are carefully examined. For each method, the convergence of the eigenvalues with
respect to the mesh refinement and the shape function order, as well as computation time and memory requirements are
investigated. The open-source template model used to obtain {\crevme the numerical results is provided}. 
Details of the implementation of polynomial and rational eigenvalue problems in GetDP are {\crevme given in appendix}.
\end{abstract}



\section{Program Summary}
\emph{Manuscript title}: Non-linear eigenvalue problems with GetDP and SLEPc: Eigenmode computations of frequency-dispersive photonic open structures.\\
\emph{Authors}: G. Dem{\'e}sy, A. Nicolet, B. Gralak, C. Geuzaine, C. Campos and J.~E. Roman\\
\emph{Program title}: NonLinearEVP.pro \\
\emph{Licensing provisions}: GNU General Public License 3 (GPL)\\
\emph{Programming language}: Gmsh (http://gmsh.info), GetDP (http://getdp.info)\\
\emph{Computer(s) for which the program has been designed}: PC, Mac, Tablets, Computer clusters\\
\emph{Operating system(s) for which the program has been designed}: Linux, Windows, MacOSX\\
\emph{Keywords}: Finite Element Method; Non-linear Eigenvalue Problem; Electromagnetism; \\
\emph{Nature of problem}: Computing the eigenvalues and eigenvectors of electromagnetic wave problems involving frequency-dispersive materials.
The resulting eigenvalue problem is non-linear and non-hermitian.\\
\emph{Solution method}: Finite element method coupled to efficient non-linear eigenvalue solvers: Relevant SLEPc solvers were 
interfaced to the Finite Element software GetDP. Several linearization schemes are benchmarked.\\
\emph{Running time}: From a few seconds for simple problems to several days for large-scale simulations.\\
\emph{References}: All appropriate references are contained in the section entitled References.

\section{Introduction}
{\ced The modes of a system are the source free solutions of the propagation equation governing the field behavior in a structured
media. They contain all the information regarding the intrinsic resonances of a given structure. In electromagnetism, when dealing
frequency-dispersive media, the Helmholtz equation appears as a non-linear eigenvalue problem through the frequency {\crevme dependence} of
the permittivities and permeabilities of the involved materials. In general, in wave physics (electromagnetism, acoustics,
elasticity...), classical EigenValue Problems (EVPs) become non-linear as soon as a material characteristic property strongly
depends on the frequency in the frequency range of interest \cite{betcke2013nlevp}. {\crevi We present a general framework to solve
non-linear EVPs suitable for a Finite Element (FE) discretization.} The implementation is based on the open-source finite element software
GetDP and the open-source library SLEPc.

The solutions of such problems may have important applications in electromagnetism at optical frequencies, where frequency
dispersion arises in bulk materials. Indeed, the permittivity of most bulk non-transparent materials, such as semiconductors and
metals, strongly depends on the excitation frequency \cite{jackson2007classical}. But frequency dispersion also comes into picture
when dealing with composites materials, or metamaterials, whose effective electromagnetic parameters derived from modern
homogenization schemes \cite{silveirinha2007metamaterial,alu2011first,liu2013causality} are frequency dependent. The accurate and
reliable computation of the modes of frequency-dispersive structures represents a great challenge for many applications in
nanophotonics.}

For smooth and monotonic material dispersion relations, it is possible to think of an iterative process where
one would set the permittivity, solve a linear EVP, adjust the permittivity value if necessary, and repeat the
process hoping for reasonable convergence for a single eigenvalue... For more tormented dispersion relations,
\emph{i.e.} in the vicinity of an intrinsic resonance of a given material, this simple iterative process is
very likely to fail. For instance, a direct determination of the spectrum of a 3D gold nanoparticle embedded
into a silicon background in the visible range is nowadays extremely challenging.

Fortunately, the relative permittivity function can be accurately described as an analytical function of the
frequency. The most famous models are the Drude, Lorentz, Debye models \cite{jackson2007classical}, the
so-called critical points \cite{etchegoin2006analytic} model or, in general, a rational function of the
frequency \cite{garcia2017extracting}. In this frame, the non linear EVP becomes rational and can be easily
transformed into a polynomial EVP.

In this paper, we numerically investigate various linearization scenarios. We apply these approaches to an
emblematic example in electromagnetism, the study of diffraction gratings. The dispersion relation of a grating
is indeed the corner stone of its physical analysis.

The recent literature on modal analysis of such open structures, referred to as Quasi-Normal Modes (QNM), is
quite rich. Even if the question of completeness and orthogonality of the QNMs remains open theoretically,
numerical quasi-normal modes expansion have been successfully used in various electromagnetic problems,
allowing to explain in an elegant manner the resonant mechanisms of a structure and its excitation condition
\cite{sauvan2013theory,vial2014optical}. Their application in nanophotonics can be found in
Refs.~\cite{sauvan2013theory,yan2018rigorous}. As described in the review article in
Ref.~\cite{lalanne2018light}, some numerical approaches already address the problem of the non-linearity of the
eigenvalue problem induced by frequency dispersion. A family of \enquote{pole search} methods
\cite{van2003band,bai2013efficient,weiss2016dark} allows to determine eigenvalues one by one by looking for
poles of a the determinant of a scattering matrix into the complex plane. Nonetheless, getting the full
spectrum in one single computation remains a harsh challenge. Given the spatial nature of the discretization
when using FE, the eigenvalue can be factorized in the final assembled matrix system. This
fundamental aspect has a {\crevme fortunate consequence: it} is possible to extract all the eigenvalues of the discrete
system in one single computation. A Finite Difference Frequency Domain (FDFD) scheme leads to the same property
and has been applied recently to open and dispersive electromagnetic structures
\cite{zimmerling2016lanczos,zimmerling2016efficient}. It relies on a square Yee grid. Finally, Boundary
Elements (BE) have been used \cite{powell2014resonant,powell2017interference} to calculate the QNMs of
dispersive arbitrarily shaped yet homogeneous structures. Since this method relies on the Green's function,
which is eigenfrequency-dependent, a contour integration has to be  \cite{powell2017interference}.

We propose to compare several FE schemes to address the non-linear EVP arising from
the frequency dispersion. The discrete problem is tackled using recent and efficient algorithms. In the last
decade, the numerical analysis community has made significant progress in the numerical solution of non-linear
eigenvalue problems, in understanding stability and conditioning issues, and also in proposing effective
algorithms. Of particular interest for this paper are iterative methods for computing a few eigenvalues and
corresponding eigenvectors of large-scale problems. {\crevme These kinds of methods} have been developed for the case of
polynomial eigenvalue problems \cite{Tisseur:2001:QEP,Mackey:2015:PEP}, but also for the more general non-linear
case \cite{Guttel:2017:NEP}. The latter includes the rational eigenvalue problem, which is indeed relevant for
the present case involving a permittivity function explicitly given as a rational function of the eigenvalue.
{\crevi These methods have proved to be effective and some of them are available in the form of robust
and efficient implementations in the SLEPc library \cite{Hernandez:2005:SSF}.} With these new solvers, one can
routinely compute selected portions of interest of the spectrum of problems with thousands of unknowns on a
mere laptop.

The paper is organized as follows. {\crevi After recalling the mathematical background we present five different 
approaches to address the non-linearity in the modal problem.}
 For each approach, a variational formulation is derived. {\ced These formulations lead to five distinct EVPs: one
rational EVP and four polynomial EVPs with various degrees (2, 3 or 4). In a second step, the corresponding discrete problems are
numerically benchmarked using the state-of-the-art SLEPc \cite{Hernandez:2005:SSF} solvers.} The issues inherent to the
sign-changing coefficients and corners are discussed and the convergence of the fundamental mode of the structure is studied. A
discussion on the respective strengths and limitations of all the proposed solutions is conducted. {\ced For the purpose of this
study, an interface to SLEPc has been implemented in the FE code GetDP \cite{getdp}. A general description of the
implementation of polynomial and rational EVPs in GetDP is given in \ref{ap:getdp}. A template open-source model showing the
implementation of each method is provided \cite{nleig_code}.}

\section{Problem statement} 
A practical challenge in computational electromagnetism is the computation, as precise and fast as possible, 
of many eigenfrequencies of a complicated 3D problem involving
frequency-dispersive permittivities and permeabilities. The photonic structure is fully described by
two space periodic tensor fields, its relative permittivity $\tensepsr(\br,\om)$ and its relative permeability
$\tensmur(\br)$, where $\br=(x,y,z)$. Note that the permeability tensor is chosen to be non dispersive
here because it is the most frequent case when dealing with bulk materials in the optical range.
The eigenvalue problem amounts to look for non trivial solutions of the source free
Maxwell's equations:

\begin{equation} \label{eq:max}
 \begin{bmatrix}
   0& \i(\epz\tensepsr(\br,\om))^{-1}\,\curl\,\cdot\\[1mm]
   -\i(\muz\tensmur(\r))^{-1}\,\curl\,\cdot&0      \\[1mm]
 \end{bmatrix} 
 \begin{bmatrix}
  \bE(\r)   \\[1mm]
  \bH(\r)   \\[1mm]
  \end{bmatrix} 
 = \om
  \begin{bmatrix}
   \bE(\r)   \\[1mm]
   \bH(\r)   \\[1mm]
   \end{bmatrix} \,.
\end{equation}
Since exploring the possible ways to linearize this problem is a complicated problem in itself, the choice is made to consider 
a structure as simple as possible and yet highlighting all the difficulties of realistic 3D structures: A
mono-dimensional grating made of frequency-dispersive rods, \emph{i.e.} a 2D structure presenting one axis of invariance
along $z$ and one direction of periodicity along $x$.  The 2D space variable is from now on denoted by $\x:=(x,y)$.

{\crevme Provided that} the constitutive tensors of materials have the form
\begin{equation}\label{eq:zaniso}
  \tensepsr=
   \begin{bmatrix}
     \varepsilon_{xx}&\varepsilon_{a}&0 \\[1mm]
     \overline{\varepsilon_a}&\varepsilon_{yy}&0\\[1mm]
  	 0&0&\varepsilon_{zz}
   \end{bmatrix}
  \mbox{ and }\tensmur=
 \begin{bmatrix}
   \mu_{xx}&\mu_{a}&0 \\[1mm]
   \overline{\mu_a}&\mu_{yy}&0\\[1mm]
	 0&0&\mu_{zz}
 \end{bmatrix},
\end{equation}
the 2D problem can be decoupled into two fundamental polarization cases. They are referred to as
$s$-pol (the electric field is along the axis of invariance) and $p$-pol (the magnetic field is along the axis of
invariance, while the electric field is orthogonal to the axis of invariance). In this paper, the choice is made to
focus on the more challenging $p$-pol case since the $s$-pol case is easier to tackle \cite{brule2016calculation}. In
particular, this polarization case {\crevme leads to surface plasmons} and it is far more representative of the difficulties
at stake in the general 3D case.

In the $p$-pol case, we denote the non vanishing electromagnetic field components by $\bH = \sh(\x)\,\bz$
and $\bE =E_x(\x)\,\bx+E_y(\x)\,\by$. The traditional choice for the unknown in the 2D $p$-pol case
is usually the out-of-plane magnetic field since the problem becomes scalar. Making use of  
$-\curl\left[\tensepsr(\x,\om)^{-1}\,\curl\,\bH \right]=
\div\left[\tensepsr(\x,\om)^{T}/\mathrm{det}(\tensepsr(\x,\om))\,\grad\,\sh \right]$, 
the resulting scalar wave equation writes in absence of electromagnetic source:
\begin{equation}\label{eq:helmholtz_h}
	-{\mu_r}_{zz}(\x)^{-1}\,\div\left[\frac{\tensepsr(\x,\om)^{T}}{\mathrm{det}(\tensepsr(\x,\om))}\,\grad\,\sh \right]
    =\frac{\om^2}{c^2}\,\sh \,.
\end{equation}
A less traditional choice for the $p$-polarization case consists in working with the in-plane electric field $\bE$ and the vector
wave equation:
\begin{equation}\label{eq:helmholtz_E}
	\tensepsr(\x,\om)^{-1}\,\curl\,\left[\tensmur^{-1}(\x)\,\curl\, \bE\right]=\frac{\om^2}{c^2}\,\bE \,.
\end{equation}
What follows is precisely meant to be extended straightforwardly to realistic 3D configurations,
where vector fields/edge elements, just as in the 2D vector case, will be at stake. As a
consequence, and even though this choice leads to larger problems at the discrete level due to the larger
connectivity of edge elements, this vector case is chosen to be the reference problem.

Note that, given the location of the dispersive permittivity in the two wave equations above, it seems more
reasonable at first glance to adopt the vector case where $\tensepsr(\x,\om)$ is outside the differential operator.
As will be shown later, one can arbitrarily choose to consider $\bE$ or $\bH$ as the unknown of the problem under
weak formulation. In fact, the scalar problem in Eq.~(\ref{eq:helmholtz_h}) will be solved as well for
enlightening comparison purposes.

The above equations constitute eigenvalue problems where $\om^2/c^2$ appears as a possible eigenvalue of
$\om$-dependent operators 
through the $\omega$ dependence of the relative permittivity. In other words, modal analysis of
frequency-dispersive structures represents a non-linear eigenvalue problem.

\section{Opto-geometric characteristics of the model}
\subsection{Geometry}
The formalism presented in this paper is very general in the sense that the tensor fields
$\tensepsr(\x,\om)$ and $\tensmur(\x)$ can be defined by part representing the distinct materials of the structure. 
Several dispersive materials can be considered and modeled by a rational function with an 
arbitrarily high number of poles. Graded-indexed and fully anisotropic materials can be handled as well.
\begin{figure}[h!]
  \centering
  \def\svgwidth{.7\textwidth}
  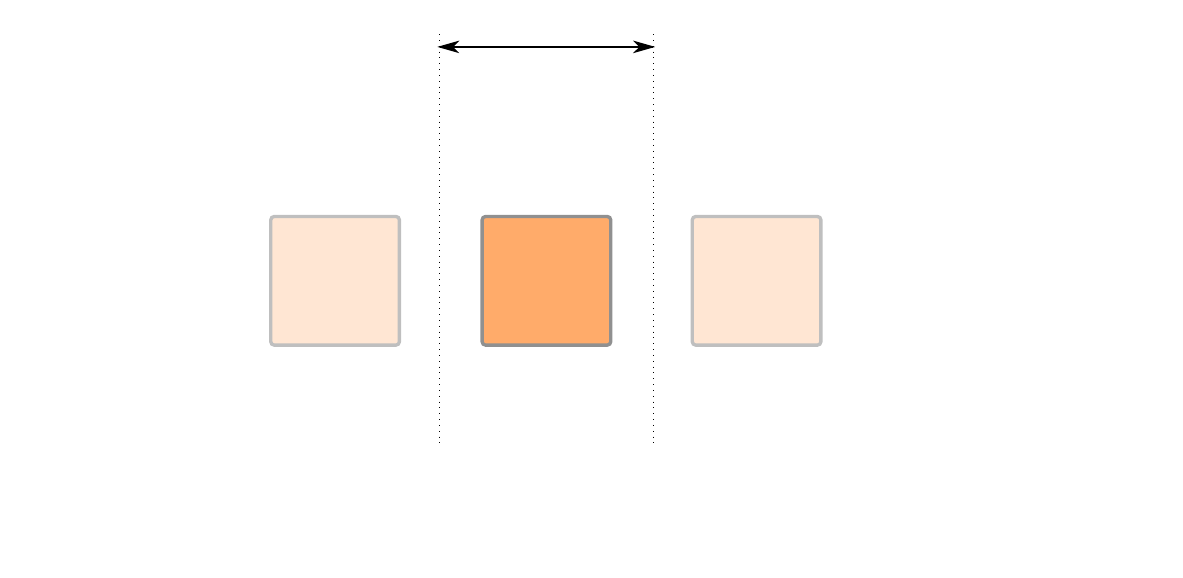 
  \caption{Geometry and notations of the problem.
  \label{fig:scheme}}
\end{figure}

In spite of the generality of the presented approach, for the sake of clarity, the derivations will be described in the
frame of the example described in Fig.~\ref{fig:scheme}. We consider from now on a simple free-standing grating with a
square section. The structure is periodic along the $x$ axis of period $a$ and invariant along the $z$-axis.
Standard cartesian Perfectly Matched Layers (PMLs \cite{teixeira1997systematic}) are used to truncate infinite extensions
of the domain along the $y$ axis. Let us denote the resulting bounded domain by $\Omega$ and its boundary by
$\partial\Omega$. The domain $\Omega$ can typically be constituted of several dispersive sub-domains with distinct
frequency-dispersion relations (in this case, one single rod with support $\Omega^{d}_1$ of boundary $\Gamma_1$) and of
several non-dispersive sub-domains. All the sub-domains ruled by the same dispersion law can be gathered together since
they can be handled all at once. Hence all non-dispersive domains are denoted by
$\Omega^{\toslash}=\bigcup_i\Omega^{\toslash}_i$. Finally, for each subset $\Omega_i$, let $\mathsf{I}_{\Omega_i}$ be its
characteristic function: $\mathsf{I}_{\Omega_i}(\x) = 1 \text{ if } \x \in \Omega_i \text{ and } \mathsf{I}_{\Omega_i}(\x)
= 0$ otherwise.

\subsection{Material properties}
The background is free-space (relative permittivity constant and equal to~1) and the rods are
made of a Drude material. Their relative permittivity $\eps{r}{1}$ writes classically \cite{jackson2007classical}:
\begin{subequations}
  \begin{align}
    \eps{r}{1}\ofom &= \epsinf - \frac{\omd^2}{\om(\om+i\gamd)} \label{eq:epsdrude1}\\
                    &= \frac{-\epsinf(i\om)^2+\epsinf\gamd(i\om)-\omd^2}{-(i\om)^2+\gamd(i\om)}  \label{eq:epsdrude3}\\
                    &:= \frac{\mathcal{N}_1(i\om)}{\mathcal{D}_1(i\om)}  \label{eq:epsdrude4}\,\, ,
  \end{align}
\end{subequations}
where $\gamd$, $\epsinf$ and $\omd$ are real constants. It is important to note that the
Drude model is causal and that $\eps{r}{1}$ is a rational function of $i\om$ with
\emph{real} coefficients (see Eq.~(\ref{eq:epsdrude3})). Finally, more realistic causal
models than the Drude model have been found and one can
generally write $\eps{r}{1}$ as a rational function (see Eq.~(\ref{eq:epsdrude4}),  where 
$\mathcal{N}_1$ and $\mathcal{D}_1$ are polynomial functions of $i\om$). It would be 
straightforward to extend this derivation to the more general case involving several frequency-dispersive domains $\Omega_i^d$
characterized by their permittivity $\eps{r}{i}\ofom$ modeled by a causal rational function:
\begin{equation}
  \eps{r}{i}\ofom = \frac{\mathcal{N}_i(i\om)}{\mathcal{D}_i(i\om)} = \frac{\sum_{j=1}^{N_i}n_{i,j}\,(i\om)^j}{\sum_{j=1}^{D_i}d_{i,j}\,(i\om)^j}\,,
\end{equation}
where $n_{i,j}$ and $d_{i,j}$ have to be real constants as detailed in Ref.~\cite{garcia2017extracting}.


Finally, the unbounded nature of the problem is handled using PMLs. The reasons for this choice {\crevme is twofold:} (i)
from the theoretical point of view, PMLs allow to reveal \cite{vial_quasimodal_2014} the so-called Quasi-Normal Modes
(PMLs can be regarded as an analytic continuation in the complex plane) and (ii) from the practical point of view,
they allow to bound the computational domain (the complex change of variable is encoded into $\tensepsr$ and
$\tensmur$ resulting in a semi-infinite layer that is eventually truncated).

Discussing the most appropriate PML parameters (\emph{i.e.} damping profile) is outside the scope of this paper,
although it would be interesting to apply many of the results obtained in time and time-harmonic domains
\cite{bermudez2007optimal,modave2014optimizing} to the eigenvalue problem. {\crevme The simplest} constant complex stretch
ruled by the complex scalar $s_y=a+ib$ is used here. The complex PML tensor is denoted then
$\tenspml=\mathrm{Diag}(s_y,1/s_y,s_y)$. One can eventually write the piecewise constant in space
and frequency-dependent relative permittivity tensor of the problem as:
\begin{equation}\label{eq:epspml}
    \tensepsr(\x,\om)=
    \begin{cases} 
       \eps{r}{1}\ofom \, \tensid & \mbox{if $\x \in \Omega_1^d$ } \\
       \eps{r}{2}      \, \tensid & \mbox{if $\x \in \Omega_2^{\toslash}$ } \\
       \eps{r}{2}      \, \tenspml & \mbox{if $\x \in \Omega_1^{\toslash}\cup \Omega_3^{\toslash}$ }
    \end{cases}
    .
\end{equation}
The piecewise constant relative permeability tensor of the problem writes :
\begin{equation}
    \tensmur(\x)=
    \begin{cases} 
        \tensid  & \mbox{if $\x \in \Omega_1^{d}\cup\Omega_2^{\toslash}$ } \\
        \tenspml & \mbox{if $\x \in \Omega_1^{\toslash} \cup \Omega_3^{\toslash}$ } 
    \end{cases}
    .
\end{equation}

Finally, Bloch-Floquet theorem is applied to the periodic structure. The problem becomes parametrized by a real
scalar $\alpha$ which spans the reduced 1D Brillouin zone $[0,\pi/a]$. In return, the study is restricted to
$(a,\alpha)$ quasi-periodic solutions (eigenvectors) of the form $\bE = \bE_\sharp\,e^{i\alpha x}$, where
$\bE_\sharp$ is a $a$-periodic vector field \cite{zolla2005foundations}. 

\subsection{Function spaces}
Several function spaces are needed to formulate the different approaches of the problem
described in the next section.

Concerning the $p$-pol vector case described in Eq.~(\ref{eq:helmholtz_E}), Bloch boundary conditions
\cite{zolla2005foundations} are applied on lateral boundaries $\Gamma_l\cup\Gamma_r$. If infinite perfectly matched layers
are the appropriate theoretical tool to reveal the quasi-normal modes by rotating the continuous spectrum into the complex
plane, they have to be truncated in practice. Truncating the PML discretizes the rotated continuous spectrum and one can
choose to apply Dirichlet or Neumann boundary conditions {\crevme at the bottom of the PML}, resulting in a slightly different
discretization as detailed in Ref.~\cite{vial_quasimodal_2014}. We choose homogeneous Dirichlet conditions on
$\Gamma_b\cup\Gamma_t$ which decreases the number of unknowns. Let us define the following Sobolev space of $(a,\alpha)$
quasi-periodic vector fields vanishing on $\Gamma_b\cup\Gamma_t$:

    \begin{multline}\label{eq:fspace}
        \bm{\mathcal{H}}_{\alpha,0}(\Omega,\curl) =\Big\{ \bE\in \left(L^2\right)^2 : \curl\,\bE\in\left(L^2\right)^2 , \\
                 \bE=e^{\i \alpha a} \bE \mbox{ and }
                 \bn_{|_{\Gamma_b}}\times\bE = 
                 \bn_{|_{\Gamma_t}}\times\bE=\mathbf{0}\Big\} \, .
     \end{multline}

The same considerations apply to the $p$-pol scalar case described in Eq.~(\ref{eq:helmholtz_h}).
However, in order to keep the same discretization of the continuous spectrum, we apply homogeneous
Neumann conditions on $\Gamma_b\cup\Gamma_t$. Let us define the following Sobolev
space of $(a,\alpha)$ quasi-periodic scalar fields:
\begin{multline}\label{eq:fspace_grad}
    \bm{\mathcal{H}}_\alpha(\Omega,\grad) =\Big\{ \sh\in L^2 : \grad\,\sh\in{\crevme\left(L^2\right)^2} ,\\
          \sh_{|_{\Gamma_r}}=e^{\i \alpha a}\sh_{|_{\Gamma_l}} \mbox{ and }
           \grad\,\sh\cdot\bn_{|_{\Gamma_b}}= 
                  \grad\,\sh\cdot\bn_{|_{\Gamma_t}}=0\Big\} \, .
\end{multline}

\section{Dealing with the eigenvalue problem non-linearity}
\subsection{A physical linearization via auxiliary fields (Aux-E case)}\label{sec:Aux-E}
The problem is reformulated using auxiliary physical fields
\cite{tip1998linear,gralak2010macroscopic}, as detailed in our previous work in
Ref.~\cite{brule2016calculation}. The procedure to obtain this extension of the Maxwell's
classical operator is briefly recalled here. By defining an auxiliary field \cite{tip2006} for
each resonance (pole) of the permittivity that couples with classical electromagnetic fields,
one can extend and linearize the classical Maxwell operator. In the present case of a simple
Drude model recalled in Eq.~(\ref{eq:epsdrude1}), a single auxiliary field denoted $\AD$ is
required, and defined in frequency-domain as:
\begin{equation} 
  \AD(\x,t) = -2\i\dfrac{\om_d}{\sqrt{2}}\dint_{-\infty}^{t}\exp[-\gamma_d(t-s)]\,\bE(\x,s)\,\mathrm{d}s .
\label{AuxD}
\end{equation}
This auxiliary field $\AD$ has for spatial support $\Omega^d_1$ and satisfies natural
boundary conditions on $\Gamma_1$. It belongs to $\bm{\mathcal{H}}(\Omega_1^d,\curl)$. An
intermediate frequency-dispersion free permittivity tensor field $\tensepsr^{\toslash}(\x)$ is
convenient here:
\begin{equation}
    \tensepsr^{\toslash}(\x)=
    \begin{cases} 
       \epsinf    \, \tensid & \mbox{if $\x \in \Omega_1^d$ } \\
       \eps{r}{2} \, \tensid & \mbox{if $\x \in \Omega_2^{\toslash}$ } \\
       \eps{r}{2} \, \tenspml & \mbox{if $\x \in \Omega_1^{\toslash}\cup \Omega_3^{\toslash}$ }
    \end{cases}
    .
\end{equation}
In matrix form, the following linear eigenvalue problem is obtained:
\begin{equation}
  \M(\x)\,\U(\x)= \om\,\U(\x)\, ,
\end{equation}
where $\U(\x)=[\bE(\x),\bH(\x),\AD(\x)]^T$ and
\begin{equation} \label{eq:evp_aux}
 \M(\x)=
 \begin{bmatrix}
   0& \i(\epz\tensepsr^{\toslash})^{-1}\,\curl\,\cdot&\dfrac{\om_d}{\sqrt{2}}{\tensepsr^{\toslash}}^{-1}\\[1mm]
   -\i(\muz\tensmur)^{-1}\,\curl\,\cdot&0&0      \\[1mm]
	 2\dfrac{\om_d}{\sqrt{2}}&0&-\i\gamma_d\\[1mm]
 \end{bmatrix} .
\end{equation}
Note that when discretizing the problem using FE, the electric field and magnetic field
cannot be represented on the same edges. The former should be discretized on the dual basis of the
latter. However, the basis functions associated with the dual unstructured FEM mesh are not easy to
construct. A possible workaround would consist in working with face elements and the 2-form
$\mathbf{B}$ instead of edge elements and the 1-form $\mathbf{H}$. Alternatively, in this paper, we
classically chose to eliminate $\bH$. The cost is that a quadratic eigenproblem is obtained whereas
the system in Eq.~(\ref{eq:evp_aux}) was linear:
\begin{equation} \label{eq:qep_aux} 
  \om^2\,\M_2(\x)\,\mathcal{V}(\x)+\om\,\M_1(\x)\,\mathcal{V}(\x)+\M_0(\x)\,\mathcal{V}(\x)=\mathbf{0}\,,
\end{equation}
where $\mathcal{V}(\x)= [\bE(\x),\AD(\x)]^T$ and
\begin{equation} \label{eq:qep_aux_mat}
   \M_2 = 
     \begin{bmatrix}
       -\tensepsr^{\toslash} & 0 \\[2mm]
       0 & 0
     \end{bmatrix}\mbox{,  }
   \M_1 = 
     \begin{bmatrix}
       0 & \dfrac{\om_d}{\sqrt{2}} \\[2mm]
       0 & -1
     \end{bmatrix}\mbox{,  }
   \M_0 =
     \begin{bmatrix}
       c^2\,\curl\left[\tensmur^{-1}\,\curl\,\cdot \right] & 0\\[2mm]
       2\dfrac{\om_d}{\sqrt{2}} & -\i\gamma_d
     \end{bmatrix}.
\end{equation}
Finally, this quadratic eigenvalue problem writes under variational form:
\begin{equation}\label{eq:Aux-E}
  \left|
  \begin{aligned}
    &  \mbox{Given $\alpha\in[0,\pi/a]$, }\\
    &  \mbox{find $(\om,\left[\bE,\AD\right]^T)
            \in\mathbb{C}\times\left[
            \bm{\mathcal{H}}_{\alpha,0}(\Omega,\curl)\times\bm{\mathcal{H}}(\Omega_1^d,\curl)
            \right]$ such that:} \\[2mm]
    & \quad \forall\mathcal{W}=\begin{bmatrix}\bW\,,\,\bW_a\end{bmatrix}
         \in \bm{\mathcal{H}}_{\alpha,0}(\Omega,\curl)\times\bm{\mathcal{H}}(\Omega_1^d,\curl),\\
    & \quad\quad \om^2\int_\Omega(\M_2\mathcal{V})\,\overline{\mathcal{W}}\,\mathrm{d}\Omega +
                 \om  \int_\Omega(\M_1\mathcal{V})\,\overline{\mathcal{W}}\,\mathrm{d}\Omega +
                      \int_\Omega(\M_0\mathcal{V})\,\overline{\mathcal{W}}\,\mathrm{d}\Omega = 0.
  \end{aligned}  
  \right.
\end{equation}
This linearization can be described as a physical one since, unlike the purely numerical ones in the
following, a larger system is obtained with extra unknowns \emph{inside the dispersive element solely}.
In this simplified version of the auxiliary fields theory called the resonance formalism, the auxiliary
field fulfills a simple relation with the polarization vector: $\partial_t
\mathbf{P}(\x,t)=\i\epz\frac{\om_d}{\sqrt{2}}\AD(\x,t)$. This approach is identical to the one presented
by Fan \emph{et al.} in Ref.~\cite{raman2010photonic}. It is also very similar to the treatment of
frequency-dispersive media made in time domain methods for direct problems such as FDTD
\cite{taflove2005computational}.

In the following, the case described in Eq.~(\ref{eq:Aux-E}) will be referred to as the Aux-E case.

\subsection{Electric field polynomial eigenvalue problem (PEP-E and NEP-E cases)}
\label{sec:PEP-E}
In this section, a purely numerical linearization is considered. This approach begins with writing
the eigenvalue problem Eq.~(\ref{eq:helmholtz_E}) under its variational form:
\begin{equation}
  \left|
  \begin{aligned}
    &  \mbox{Given $\alpha\in[0,\pi/a]$, find $(\om,\bE)\in\mathbb{C}\times\bm{\mathcal{H}}_{\alpha,0}(\Omega,\curl)$ such that:} \\
    & \quad \forall\,\bW\in\bm{\mathcal{H}}_{\alpha,0}(\Omega,\curl),\\
    & \quad\quad -\Galerkin{\tensmur^{-1}\,\curl\, \bE}{\curl\, \bW}{\Omega}
             + \frac{\om^2}{c^2}\Galerkin{\tensepsr(\x,\om)\,\bE}{\bW}{\Omega}=0\,.
  \end{aligned}  
  \right.
\end{equation}
Note that the boundary term on periodic lines $\Gamma_r$ and $\Gamma_l$ vanishes due to
opposite signs of normals \cite{nicolet2004modelling}.

Then, recalling that the whole domain $\Omega$ can be split into frequency-dispersive domains
($\Omega_1^d$ solely in this simplified case) and non dispersive domains $\Omega^{\toslash}$, and
that the permittivity tensor is a constant by part tensor field of $\r$, the problem becomes :
\begin{equation}\label{eq:NEP-E}
  \left|
  \begin{aligned}
    & \mbox{Given $\alpha\in[0,\pi/a]$, find $(\om,\bE)\in\mathbb{C}\times\bm{\mathcal{H}}_{\alpha,0}(\Omega,\curl)$ such that:} \\
    & \quad \forall\,\bW\in\bm{\mathcal{H}}_{\alpha,0}(\Omega,\curl),\\
    & \quad\quad\quad        -\Galerkin{\tensmur^{-1}\,\curl\, \bE}{\curl\, \bW}{\Omega}\\
    & \quad\quad\quad +\frac{\om^2}{c^2}\Galerkin{\tensepsr^{\toslash}\,\bE}{\bW}{\Omega^{\toslash}}
                      +\frac{\om^2}{c^2}\frac{\N_1(i\omega)}{\D_1(i\omega)}\Galerkin{\bE}{\bW}{\Omega_1^d}=0\,.
  \end{aligned}  
  \right.
\end{equation}
A last mere multiplication by $\D_1(i\omega)$ allows to express the problem under the form of
a polynomial eigenvalue problem:
\begin{equation}\label{eq:PEP-E}
  \left|
  \begin{aligned}
    &  \mbox{Given $\alpha\in[0,\pi/a]$, find $(\om,\bE)\in\mathbb{C}\times\bm{\mathcal{H}}_{\alpha,0}(\Omega,\curl)$ such that:} \\
    & \quad \forall\,\bW\in\bm{\mathcal{H}}_{\alpha,0}(\Omega,\curl),\\
    & \quad\quad\quad        -\D_1(i\omega)\Galerkin{\tensmur^{-1}\,\curl\, \bE}{\curl\, \bW}{\Omega}\\
    & \quad\quad\quad +\frac{\om^2}{c^2}\D_1(i\omega)\Galerkin{\tensepsr^{\toslash}\,\bE}{\bW}{\Omega^{\toslash}}
                      +\frac{\om^2}{c^2}\N_1(i\omega)\Galerkin{\bE}{\bW}{\Omega_1^d}=0.
  \end{aligned}  
  \right.
\end{equation}
The Drude permittivity model has a pole in zero, leading to a polynomial EVP of order $3$.
Otherwise, when considering one single frequency-dispersive material, the final order will be
$2+\mathrm{Deg}(\D_i)$. More generally, note that the final degree of the polynomial EVP is
$2+\sum_{i=1}^{N} \mathrm{Deg}(\D_i)$ in the case of $N$ (distinct) frequency-dispersive materials.

In the following, the approaches described in Eq.~(\ref{eq:PEP-E}) and Eq.~(\ref{eq:NEP-E})
will be referred to as the PEP-E and NEP-E approaches respectively (resp.). They differ by the type 
of solver used for their numerical treatment as detailed later.

\subsection{Electric field polynomial eigenvalue problem with Lagrange multipliers (Lag-E case)} 
One can consider the polynomial eigenvalue problem
under its strong form, by a mere multiplication of the propagation equation by the denominator of the
frequency-dispersive permittivity. Recalling that the relative permittivity tensor field is defined by 
part in each domain, we obtain:
\begin{multline}\label{eq:pep_strong}
  \left(\mathsf{I}_{\Omega^\toslash}(\x)+\mathcal{D}_1(i\om)\,\mathsf{I}_{\Omega^d_1}(\x)\right)
  \curl\,\left[\tensmur^{-1}(\x)\,\curl\, \bE\right]=\\
  \left(\tensepsr(\x)\,\mathsf{I}_{\Omega^\toslash}+
  \mathcal{N}_1(i\om)\mathsf{I}_{\Omega^d_1}(\x)\right)\frac{\om^2}{c^2}\,\bE \,.
\end{multline} 
Terms of the form $\left[ f(\x)\,\curl\, \tensmur^{-1} \curl\, \bE\right]$ are obtained, where $f$
is a constant by part complex scalar function. The weak formulation is not classical, since after
multiplication by a test function $\bW$ and integration over $\Omega$, we obtain in the sense of
distributions :
\begin{equation}\label{eq:intextra}
  \begin{split}
   \displaystyle \int_\Omega \left[ f(\bx)\,\curl\,\tensmur^{-1} \curl\, \bE\right]\cdot\overline{\bW}\,\mathrm{d}\Omega &
   =   \displaystyle \int_\Omega f(\bx)\,\tensmur^{-1}\,
       \curl\,\bE\cdot\curl\,\overline{\bW}\,\mathrm{d}\Omega \\
   & - \displaystyle \int_{\partial\Omega} f(\bx)\,\left[ \tensmur^{-1}\,
       \curl\,\bE\times\bn_{|_{\partial\Omega}}\right]\cdot\overline{\bW}\,\mathrm{d}\Gamma \\
   & + \displaystyle \int_{\Gamma_1} f_{\mathrm{jump}}^{\toslash\rightarrow d}
       \left[\left[ \tensmur^{-1}\,\curl\,  \bE\right]\times\bn_{|_{\Gamma_1}}\right]\cdot \overline{\bW}\,\mathrm{d}\Gamma \,,
  \end{split} 
\end{equation}
where $f_{\mathrm{jump}}^{\toslash\rightarrow d}$ is the jump of $f$ across $\Gamma_1$.
The two first terms in the right hand side of Eq.~(\ref{eq:intextra}) are exactly like those arising from the traditional
integration by part of the $\curl$ (pondered by $f$). As for the last term, it represents a jump to enforce the
quantity $\left[\left[ \tensmur^{-1}\,\curl\, \bE\right]\times\bn_{|_{\Gamma_1}}\right]$, which is nothing but the tangential
trace of $\tensmur^{-1}\,\curl\, \bE$ on $\Gamma_1$. This quantity is not readily accessible and requires the adjunction of a
Lagrange multiplier. In other words, the procedure now consists in splitting the problem into groups ruled by the same
frequency dispersion law and introducing an extra unknown in order to reassemble the different groups while satisfying the
appropriate fields discontinuities. Thus, the problem is split into two distinct parts and two fields $\bE_1$ and $\bE_2$ are
defined, with respective support $\Omega_1^d$ and $\Omega^\toslash$. A Lagrange multiplier $\bm{\lambda}$ is introduced on
$\Gamma_1$ in order to set the appropriate boundary conditions. It remains to define the appropriate trace space of
$\Omega_1^d$ on $\Gamma_1$ \cite{monk2003finite}:
$\bm{\mathcal{H}^{-1/2}}(\div,\Gamma_1)=\{\mathbf{u}\times\bn_{|_{\Gamma_1}}: \mathbf{u}\in\mathcal{H}(\curl,\Omega_1) \}$
which coincides with the trace space of $\Omega^\toslash$ on $\Gamma_1$ up to the orientation of the normals.

The variational form of the eigenproblem writes:

\begin{subequations}\label{eq:Lag-E}
\begin{empheq}[left=\empheqlvert]{align}
  & \mbox{Given $\alpha\in[0,\pi/a]$, find $(\om,(\bE_1,\bE_2,\bm{\lambda})) \in$} \nonumber\\
  & \mbox{$\mathbb{C}\times\left[
    \bm{\mathcal{H}}(\Omega^d_1,\curl)
    \times \bm{\mathcal{H}}_{\alpha,0}(\Omega^\toslash,\curl)
    \times \bm{\mathcal{H}^{-1/2}}(\div,\Gamma_1)
  \right]$ such that:}  \nonumber \\
  & \quad \forall\,\left[\bW_1,\bW_2,\bm{\nu}\right]^T\in \nonumber\\
  & \quad\quad\bm{\mathcal{H}}(\Omega^d_1,\curl)
    \times\bm{\mathcal{H}}_{\alpha,0}(\Omega^\toslash,\curl)
    \times{ \bm{\mathcal{H}^{-1/2}}(\div,\Gamma_1)}, \nonumber \\[2mm]
  & \quad\quad\quad \bullet \mathcal{D}_1(i\om)\,\Galerkin{\tensmur^{-1}\curl\,\bE_1}{\curl\,\mathbf{W}_1}{\Omega^d_1} \nonumber \\
  &\quad\quad\quad\quad+\frac{\om^2}{c^2}\,\mathcal{N}_1(i\om)\,\Galerkin{\bE_1}{\mathbf{W}_1}{\Omega^d_1}
                  +\mathcal{D}_1(i\om)\,\Galerkinl{\boldsymbol{\lambda}}{\mathbf{W}_1}{\Gamma_1}=0 \label{seq:Lag-E_a}  \\
  & \quad\quad\quad \bullet \Galerkin{\tensmur^{-1}\curl\,\bE_2}{\curl\,\mathbf{W}_2}{ \Omega^{\toslash}} \nonumber \\
  & \quad\quad\quad\quad +\frac{\om^2}{c^2}\Galerkin{\tensepsr^{\toslash}\,\bE_2}{\mathbf{W}_2}{ \Omega^{\toslash}}
                    -\Galerkinl{\boldsymbol{\lambda}}{\mathbf{W}_2}{\Gamma_1}=0 \label{seq:Lag-E_b} \\
  &\quad\quad\quad \bullet \Galerkinl{{ \bn_{|_{\Gamma_1}}\times}(\bE_1+\bE_2)}{\boldsymbol{\nu}}{\Gamma_1}=0 \,. \label{seq:Lag-E_c}
\end{empheq}
\end{subequations}
In the system Eqs.~(\ref{eq:Lag-E}), the two first equations Eqs.~(\ref{seq:Lag-E_a},
\ref{seq:Lag-E_b}), apart from their respective last boundary term, are nothing but the
variational form of the wave equation in the dispersive domain $\Omega_1^d$
(Eq.~(\ref{seq:Lag-E_a})) and in the non-dispersive domain $\Omega^\toslash$. As for this last
boundary term, it accounts for the discontinuity of the denominator of the permittivity over
$\Omega$ through the Lagrange multiplier $\bm{\lambda}$ by imposing appropriate jumps to the
tangential trace of $\tensmur^{-1}\,\curl\, \bE$ on $\Gamma_1$. Finally, the continuity of the
tangential component of $\bE=\bE_1+\bE_2$ on $\Gamma_1$ is restored in Eq.~(\ref{seq:Lag-E_c}).

The advantage of this approach is that, in case of several dispersive materials, the degree of the final polynomial EVP
remains $\mathrm{Max}_i\{\mathrm{Deg}(\mathcal{D}_i),2+\mathrm{Deg}(\mathcal{N}_i)\}$ instead of being the sum of the degrees
the $\mathcal{D}_i$ polynomials as in Sec.~\ref{sec:PEP-E}. However, in this example where a Drude material is in contact
with a dispersion-free region, it results in a 3$^{rd}$ order polynomial as in the PEP-E approach. Note that one drawback is
the additional surface unknowns introduced by the Lagrange multipliers.

In the following, the approach described by Eqs.~(\ref{seq:Lag-E_a},\ref{seq:Lag-E_b},\ref{seq:Lag-E_c}) will be referred to
as the Lag-E approach.

\subsection{Magnetic field polynomial eigenvalue problem (PEP-h case)}
For reference and comparison, we will also solve here the scalar problem corresponding to Eq.~(\ref{eq:helmholtz_h}). Let us
recall that homogeneous Neumann boundary conditions are imposed at the extremities of the PMLs in order to keep the same
discretization of the continuous spectrum as in the other approaches based on the electric field. This
continuous scalar problem can be tackled using nodal elements whereas the previous ones requires edge elements. The same
considerations as in the previous vector case allow to establish the eigenproblem for the scalar unknown $\sh$:
\begin{equation}\label{eq:PEP-h}
  \left|
  \begin{aligned}
    &  \mbox{Given $\alpha\in[0,\pi/a]$, find $(\om,\sh)\in\mathbb{C}\times\bm{\mathcal{H}}_{\alpha,0}(\Omega,\grad)$ such that:} \\
    & \quad \forall w\in\bm{\mathcal{H}}_{\alpha,0}(\Omega,\grad),\\
    & \quad\quad-\D_1(i\omega)\Galerkin{\grad\, \sh}{\grad\, w}{ \Omega_1^d}\\
    & \quad\quad-\N_1(i\omega)\Galerkin{\frac{\tensepsr(\x)^{T}}{\mathrm{det}(\tensepsr(\x))}\,\grad\, \sh}{\grad\, w}{\Omega^\toslash}\\
    & \quad\quad+ \frac{\om^2}{c^2}\N_1(i\omega)\Galerkin{{\mu_r}_{zz}\,\sh}{w}{\Omega}=0 \,.
  \end{aligned}
  \right.
\end{equation}
In the present grating example with a Drude material, it results in a 4$^{th}$ order polynomial EVP.

In the following, the approach described in Eq.~(\ref{eq:PEP-h}) will be referred to as the PEP-h approach.
\crevi
\section{Solving the discrete problem}
\subsection{ Discretization and summary}
\begin{figure}[h!] \centering
  \includegraphics[width=.92\textwidth,draft=\flagdraftfig]{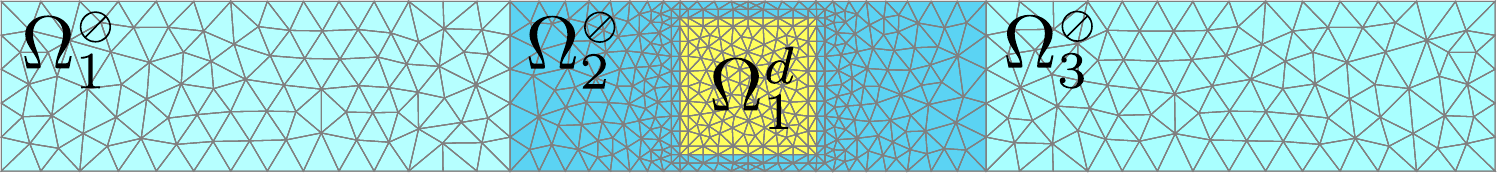}
  \caption{\crevi Mesh of the computational domain for $N=5$. The mesh size is set to $a/N$ in 
  $\Omega_2^{\toslash}$ (free-space), $a/(2N)$ in and around $\Omega_1^d$ (dispersive rod). \label{fig:mesh}}
\end{figure}

\cblack
{\crevi The structure described in Fig.~\ref{fig:scheme} was meshed using {\crevme the open source mesh generator Gmsh} \cite{gmsh}. 
A sample mesh is shown in Fig.~\ref{fig:mesh}. In the following numerical experiments, 
the distance from the object to the PML is set to $a$ and the PML thickness to $3a$. The mesh size is set to $a/N$ in
$\Omega_2^{\toslash}$ (free-space), $a/(3N)$ in and around $\Omega_1^d$ (dispersive rod), where $N$
is set to an integer value.  Note that this last value of the mesh refinement in the dispersive rod is arbitrary
since its permittivity is eigenvalue-dependent. Indeed one cannot choose the mesh size like in time-harmonic direct problems: 
In direct cases, the frequency is fixed, and thus the spatial variations characteristic length of the unknown field inside 
each domain are known in advance. As will be discussed in Sec.~\ref{sec:corners}, the mesh is globally unstructured and locally
structured in order to be symmetric at the interface with the dispersive material. Finally, the mesh is periodic in the 
direction of periodicity of the grating. }

{\crevi First or second order edge elements (or Webb elements with interpolation order $k=1$ or 2 
\cite{webb1993hierarchal,geuzaine1999convergence}) are used in electric field cases (Aux-E, PEP-E, NEP-E, Lag-E) and first or 
second order nodal elements are used in the magnetic field case (PEP-h) depending on the study. The GetDP \cite{getdp} software 
allows to handle the various required basis functions handily (details about the implementation are given in \ref{ap:getdp}). 
Finally, ONELAB is an open-source software bundle \cite{onelab}, containing both Gmsh and GetDP, which provides a lightweight 
graphical interface to these programs. A ONELAB open-source model can be downloaded from \cite{nleig_code} and allows to 
reproduce the results presented in Sec.~\ref{sec:converg}.

The different cases and their main differences (unknown field, polynomial orders, number of DOFs for a particular mesh, solver
used) are summed up in Table~\ref{tab:sumup}.}

\begin{table}[h]\footnotesize\crevme
  \centering 
  \caption{A synthetic view of all the presented approaches.}
  \label{tab:sumup}
  \small
  \begin{tabular}{|l|c|c|c|c|c|}
    \hline
    Name                 & Aux-E                & PEP-E                & NEP-E                & Lag-E                           & PEP-h                \\\hline
    Formulation          & Eq.~(\ref{eq:Aux-E}) & Eq.~(\ref{eq:PEP-E}) & Eq.~(\ref{eq:NEP-E}) & Eq.~(\ref{eq:Lag-E})            & Eq.~(\ref{eq:PEP-h}) \\
    Unknown(s)           & $\bE$, $\AD$         & $\bE$                & $\bE$                & $\bE_1$, $\bE_2$, $\bm{\lambda}$& $\sh$                \\
    Element type         & Edge                 & Edge                 & Edge                 & Edge                            & Nodal                \\
    Polynomial order     & 2                    & 3                    & (rational)           & 3                               & 4                    \\
    SLEPc solver         & PEP                  & PEP                  & NEP                  & PEP                             & PEP                  \\\hline
    Number of DOFs       & 44596                & 34150                & 34150                & 34750                           & 23688                \\
    \hline
  \end{tabular}
\end{table}

\cblack

\subsection{Solvers}
Very recent progress in sparse matrix eigenvalue solvers allow to tackle the discrete
problem very efficiently. For the purpose of this study, we interfaced GetDP with two
particularly well suited and recent solvers of the SLEPc library
\cite{Hernandez:2005:SSF} dedicated to solve large scale sparse eigenvalue problems.
Depending on the eigenproblem, GetDP can call linear, quadratic, general polynomial, or
rational eigenvalue solvers of SLEPc\footnote{A version of SLEPc 3.8.0 or more recent is required.}. 

Concerning the auxiliary field (Aux-E) formulation, all is needed is a solver adapted to
quadratic eigenproblems. Again, that is the particularity of this physical linearization,
one can add more poles to the permittivity rational function or more dispersive
materials: It will only result in defining new auxiliary fields in the elements leading
to a larger system that will remain quadratic. 

As for the polynomial eigenproblems (PEP-E, Lag-E, PEP-h) described in
Eqs.~(\ref{eq:PEP-E},\ref{eq:Lag-E},\ref{eq:PEP-h}), the matrices corresponding to the
various powers of $\omega$ (that is, in the reference example, 4 matrices for the electric field
formulations and 5 for the magnetic electric field formulation) are assembled separately
in GetDP and simply passed to SLEPc. SLEPc provides a PEP module for the solution of
polynomial eigenvalue problems, either quadratic or of higher degree $d$. The user can
choose among several solvers. Most of these solvers are based on linearization, meaning
that internally a linear eigenvalue problem is built somehow and solved with more
traditional linear eigensolvers. The linear eigenproblem produced by the linearization
is of dimension $d\cdot n$, where $n$ is the size of the polynomial problem. Hence, a
naive implementation of the linearization is going to require $d$ times as much memory
with respect to the linear case. The default SLEPc polynomial solver, named TOAR, is
memory-efficient because it represents the subspace basis in a compact way,
$V=(I_d\otimes U)G$, where vectors of the basis $U$ have length $n$ as opposed to length
$d\cdot n$ for vectors of $V$. The TOAR algorithm builds a Krylov subspace with this
basis structure, and it has been shown to be numerically stable \cite{Lu:2016:SAT}.
Apart from the memory savings, the method is cheaper in terms of computations compared
to operating with the explicitly formed linearization. In particular, when performing
the shift-and-invert spectral transformation for computing eigenvalues close to a given
target value in the complex plane, it is not necessary to factorize a matrix of order
$d\cdot n$ but a matrix of order $n$ instead. SLEPc's solvers also incorporate all the
necessary ingredients for making the method effective and accurate, such as scaling,
restart, eigenvalue locking, eigenvector extraction, and iterative refinement, as well
as parallel implementation. All the details can be found in \cite{Campos:2016:PKS}.

The rational eigenproblem described in Eq.~(\ref{eq:NEP-E}) is even simpler since SLEPc now has a
built-in solver class to handle complex rational functions. As a result, one can directly provide
the 3 necessary matrices corresponding to the tree terms in Eq.~(\ref{eq:NEP-E}), along with the
desired dispersive relative permittivity function. Note that for several dispersive domains with
distinct materials with a high number of poles, the product of all the involved denominators in the
polynomial approach (Eq.~(\ref{eq:PEP-E})) would be tedious to write. However, the number of terms
to write with the NEP solvers remains \enquote{two plus the number of distinct dispersive media}.
We present both these twin approaches, but, from the practical point of view, the rational NEP
solver class is clearly the best match for the purpose of this study.

SLEPc's NEP module for general non-linear eigenproblems {\crevme \cite{Campos:2018:NEP}} can be used
to compute a few eigenvalues (and corresponding eigenvectors) of any eigenproblem that is
non-linear with respect to the eigenvalue (not the eigenvector). This includes the
rational eigenvalue problem, for which SLEPc solvers provide specific support. The
problem is expressed in the form
\begin{equation}\label{eq:split}
\sum_{i=0}^{\ell-1}A_if_i(\lambda)x = 0,
\end{equation}
where $A_i$ are the matrix coefficients and $f_i(\cdot)$ are non-linear functions. Again,
SLEPc provides a collection of solvers from which the user can select the most
appropriate one. Particularly interesting are the methods based on approximation
followed by linearization. An example of such methods is the interpolation solver, that
approximates the non-linear function by the interpolation polynomial in a given interval,
and then uses the PEP module to solve the resulting polynomial eigenproblem. This
approach is available only for the case of real eigenvalues and hence cannot be applied
to this case. A similar strategy is used in the NLEIGS algorithm \cite{Guttel:2014:NCF},
that builds a rational interpolation which in turn is linearized to get a linear
eigenvalue problem. As opposed to the case of the polynomial eigenproblem, in this case
the dimension of the linearized problem is not known a priori, since the number of terms
depends on the function being interpolated. NLEIGS determines the number of terms based
on a tolerance for interpolation. In a general non-linear function, the user must provide
a discretization of the singularity set, but in the case that the non-linear eigenproblem
is itself rational, this is not necessary and SLEPc automatically builds an exact
rational interpolation of size equal to the number of poles (plus the degree of the polynomial part if
present). Once the rational interpolation is obtained, the last step is to create a
memory-efficient Krylov expansion associated with the linearization, in a similar way as
in polynomial problems, without explicitly building the matrix of the linearization and
representing the Krylov basis in a compact way. This is the approach that has been used
in this paper for the NEP-E formulation.
\crevi
\section{Spectrum of the dispersive grating}\label{sec:results}
\cblack
{\crevi The numerical values used in Refs.~\cite{van2003band,brule2016calculation} in the case of 2D photonic 
crystals are considered here:
\begin{equation}\label{eq:numval}
  \epsinf=1,\mbox{  }\gamd=0.05\eta,\mbox{ and } \omd =1.1\eta ,\mbox{ with }\eta=\frac{2\pi c}{a}.
\end{equation}
The square rod section is set to $b=0.806a$. For the spectra computed in this section, second order FE shape 
functions are used and the mesh parameter is set to $N=8$ (cf. Fig.~\ref{fig:mesh}).
\subsection{Dispersion relation in the complex plane}\label{sec:dispcplane}

For each formulation, the reduced Brillouin zone $[0,\pi/a]$ is sampled by 60 points. 
For each value of the Bloch variable $\alpha$, 500 complex eigenvalues are computed inside a predefined rectangular region of 
interest in the lower right quarter of the complex plane.
}
\begin{figure}[h!] \centering
  \includegraphics[width=.92\textwidth,draft=\flagdraftfig]{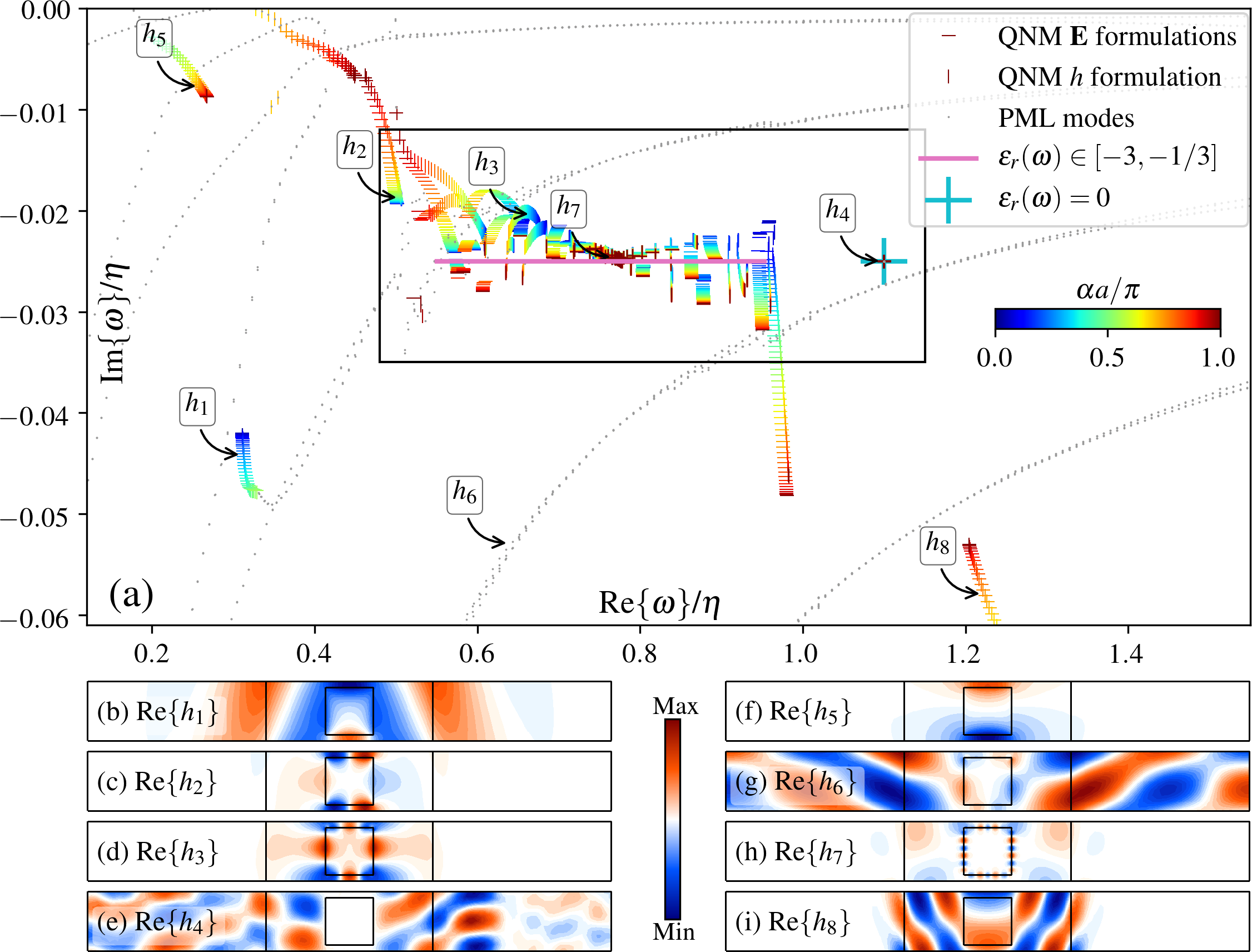}
  \caption{\crevi Dispersion relation of the grating in the complex frequency plane. 
  The reduced Brillouin zone is spanned by $\alpha$ in color scale from 0 to $\pi/a$ with 60 points. 
  All electric field formulations (Aux-E, PEP-E, NEP-E, Lag-E) 
  give identical eigenvalues up to the solver tolerance and are represented by hyphens \enquote{{\tt\_}}. The magnetic field
  formulation PEP-h is represented by vertical hyphens \enquote{{\tt|}}. The critical interval of complex frequencies due 
  to sharp corners is represented by the pink solid line. The large blue plus represents the relevant zero of $\eps{r}{1}$.}
  \label{fig:eigs_cplane}    
\end{figure}

{\crevi This set of eigenvalues forms the dispersion relation 
of the grating for which a standard representation ($\omega$ against $\alpha$) is given in \ref{ap:disp}.
Choosing a representation in the complex frequency plane brings an enlightening viewpoint. 
Indeed, the lower right quarter of the complex plane of interest (positive real part, negative imaginary part corresponding
to a damping in time) exhibits several very particular and unavoidable points.
The dispersion relations for all five problems are represented in the complex plane in Fig.~\ref{fig:eigs_cplane}(a). 
The parameter in color scale is the Bloch variable $\alpha$. Horizontal \enquote{{\tt \_}} 
and vertical \enquote{{\tt |}} hyphens represent the eigenvalues of the QNMs of interest for the electric and magnetic 
field formulations respectively. A selection of eigenvectors computed using the PEP-h formulation is given in 
Figs.~\ref{fig:eigs_cplane}(b-i).

As detailed in \ref{ap:disp}, the PML modes (represented here by grey dots, with eigenfields confined in the PMLs, see $h_6$ in 
Fig.~\ref{fig:eigs_cplane}(g)) can be numerically distinguished from QNMs using a 
twofold criterion. It is stressed that each colored curve in the complex plane is a photonic band of the grating. 
For more physical considerations about these bands, one can refer to \cite{lalanne2019quasinormal}. 
The most commonly used are the low frequency ones (close to the real line as $h_5$ or $h_2$ in 
Figs.~\ref{fig:eigs_cplane}(f,c)).

The first striking point is the perfect numerical agreement between all the approaches based on the electric field 
(colored horizontal bars \enquote{{\tt \_}} ). It is as good as the order of magnitude of the tolerance of solver which was set to $10^{-9}$ 
\cite{slepc-users-manual}. In other words, {\it all the linearization schemes presented for the electric field are numerically equivalent}.
The magnetic field formulation PEP-h fits the other cases with good accuracy outside the black rectangular frame.

Now let us look into the differences between the electric and magnetic field formulations, 
wherever horizontal and vertical bars do not form a \enquote{{\tt +}}) by recalling the properties 
of three particular regions \cite{brule2016calculation} of the complex plane: $\omega$ such that 
$\eps{r}{1}(\omega)\in[-I_\theta,-1/I_\theta]$,
$\eps{r}{1}(\omega)=0$, and
$\eps{r}{1}(\omega)\rightarrow\infty$.}

\crevi
\subsection{Corner modes and surface plasmons: $\eps{r}{1}(\omega)\in[-I_\theta,-1/I_\theta]$}\label{sec:corners}
In recent works, variational formulations of the Helmholtz equation with sign changing coefficients has drawn 
a lot of attention in both direct \cite{chesnel2013t,Bonnet-BenDhia2018} and spectral problems \cite{carvalho2017eigenvalue}. The sesquilinear form
involving the sign-changing coefficient becomes non coercive and one cannot use the Lax-Milgram theorem
to establish well-posedness. In the direct problem, with a real and fixed frequency, the problem exists
but it is hidden by the simple fact that most of physical problems are dissipative (\emph{i.e.} the
real-part changing coefficient has a non vanishing imaginary part). However, in spectral problems with
\emph{complex frequencies}, one can always find regions of the complex plane of frequencies for which the
\emph{sign-changing coefficient is purely real}.

One important starting point is that it is possible to foresee \cite{carvalho2017eigenvalue} the critical
complex frequencies: Given a sharp angle $\theta$ of the considered object with negative
permittivity, the problem is ill-posed for:
\begin{equation}
  \eps{r}{1}\in[-I_\theta,-1/I_\theta] \mbox{, where } 
  I_\theta=\mbox{max}\left(\frac{2\pi-\theta}{\theta},\frac{\theta}{2\pi-\theta}\right)\,.
\end{equation}

In this case, singularities appear at the corners and the expected corner modes are becoming more and
more oscillating in the close vicinity of the corner. These solutions are no longer of finite energy, so
in the functional frame of classical Galerkin FE used here, these modes known as \enquote{black-hole
waves} cannot be represented. It is interesting to note that corner modes correspond to continuous
spectrum just as free-space. This corner problem is a generalization of the well known one occurring at 
flat interfaces ($\theta=\pi$). The critical interval reduces then to the singleton $\{-1\}$. 

Following the meshing prescriptions in Refs.~{\cite{chesnel2013t,Bonnet-BenDhia2018}}, 
{\it a structured symmetric mesh is imposed at each sign-changing interface and corner} 
(cf. \Fig{fig:mesh}, \Fig{fig:conv}(d) and \Fig{fig:flat}). As clearly illustrated in \ref{ap:round} 
(cf. \Fig{fig:flat}) in the case of a simple planar slab, using an arbitrary unstructured mesh around a sign-changing
interface leads to a highly unstable numerical discretization of the plasmonic accumulation point.

For the present dispersive rod consisting of $\pi$/2 angles, the critical interval is $[-3,-1/3]$.
When applying the Drude model to complex frequencies, the quadratic equation $\eps{r}{1}(\om_c)=\kappa$ 
has one root $\om_c$ in the quarter complex plane of interest for any $\kappa\in[-3,-1/3]$: 
$\om_c=-i\gamd/2+\sqrt{-\gamd^2/4+\omd^2/(1-\kappa)}$. 

The thick purple segment in Figs.~\ref{fig:eigs_cplane} shows the locus of $\om_c$ as $\kappa$ spans
$[-3,-1/3]$. In other words, all the eigenvalues around this segment correspond to corner modes.
This explains the shift between the edge-based (electric) discretizations and the
nodal (magnetic) one: {\it They both fail at capturing the corner effect in a different manner}. Indeed, in the
(in-plane) edge case, the relevant unknowns associated with the corner are the circulation of the field
along the two adjacent edges discretizing the corner, whereas in the (out-of-plane) nodal case, one
unknown lies exactly on the corner. As moving closer to the critical interval, one finds higher 
order surface modes with rapid spatial frequencies (see eigenvector $h_2$, $h_3$ and finally $h_7$ in 
Fig.~\ref{fig:eigs_cplane}(c,d,h)), to finally find in the vicinity of $\omega/\eps{r}{1}(\omega)=-1$ 
undersampled eigenvectors appearing as four weighted hot spots around each corner of the square.

An interesting and rigorous workaround could consist in using a special kind of PML dedicated to 
corners \cite{carvalho2015etude}. To the best of our knowledge, this type of PML has never been implemented
to compute spectra of dispersive and lossy materials and it is a separate subject of study. 
More pragmatically, it is tempting to reduce the critical interval 
by slightly rounding the corners. In \ref{ap:round}, the  effect of a rounding on only one mesh element is shown to be quite spectacular. 
Even with rounded corners, an accumulation point of eigenvalues \cite{brule2016calculation} cannot be avoided at the 
plasmon frequency such that $\eps{r}{1}(\om)=-1$. It corresponds to all surface plasmon modes supported around the rod, 
with spatial variations tending towards infinity (cf. \ref{ap:round}).

Note that the spectrum of the original square structure is modified by both the corner PML 
(by selecting an outgoing wave condition at the corner) and the rounded geometry (obviously).

\subsection{Divergence failure: $\eps{r}{1}(\omega)=0$}\label{sec:divfail}
The second type of particular point corresponds to the zeros of the dispersive permittivity $\eps{r}{1}(\omega)$. 
With the Drude model, the region of interest exhibits a single zero shown in
Fig.~\ref{fig:eigs_cplane}(a) by a large blue \enquote{{\tt +}}. When reaching a zero of
$\eps{r}{1}$, the divergence condition $\div(\tensepsr\,\bE)=0$ fails to give information about the electric field $\bE$ 
which acquires supplemental degrees of freedom. For the 60 EVPs solved to compute the 
dispersion relation, an average number of 130 eigenvalues out of the 500 computed correspond to zeros of the permittivity.
It is of course a limitation in terms of computation time. {\crevme Note that these points are trivial} to compute so that a numerical 
workaround would consist in adding some exclusion regions of the complex plane thanks to the SLEPc region class. 
This problem does not occur in the $s$-pol case, where the only unknown is $E_z$, since 2D nodal elements
are divergence free by construction ($\partial_x E_x=\partial_y E_y=\partial_z E_z = 0$). A reason for solving 
the $p$-pol case using the PEP-h was to check whether the impact of this problem could be reduced using the magnetic field. 
It is not the case, since the average number of eigenvalues found at $\omega/\eps{r}{1}=0$ is about 130 (out of 500) for 
electric field formulations and 135 for PEP-h. Looking at the particular eigenvector labelled $h_4$ in 
Fig.~\ref{fig:eigs_cplane}(e), the field appears constant inside the dispersive rod and exhibits random fluctuations outside.
One possible workaround consists in enforcing the appropriate divergence condition by adjunction of a Lagrange multiplier.

\subsection{Bulk accumulation point: $\eps{r}{1}(\omega)\rightarrow\infty$}\label{sec:divfail}
The eigenvectors associated with eigenvalues found around the poles of the eigenvalue-dependent permittivity 
{\crevii present} no characteristic spatial variations since they can be arbitrarily rapid. For a Drude model, 
poles are 0 and $-i\gamma$. It was chosen to avoid these points using the rectangular region provided by SLEPc.
Indeed, eigenfrequencies with null real parts are usually not relevant. However, when considering a permittivity model
with poles in the region of interest ({\it e.g.} a Lorentz model), it is of prime importance to take this 
accumulation point into account as we already evidenced in \cite{zolla2018photonics}.   

\subsection{Convergence}\label{sec:converg}

\begin{figure}[h]
  \centering
  \includegraphics[width=.92\textwidth,draft=\flagdraftfig]{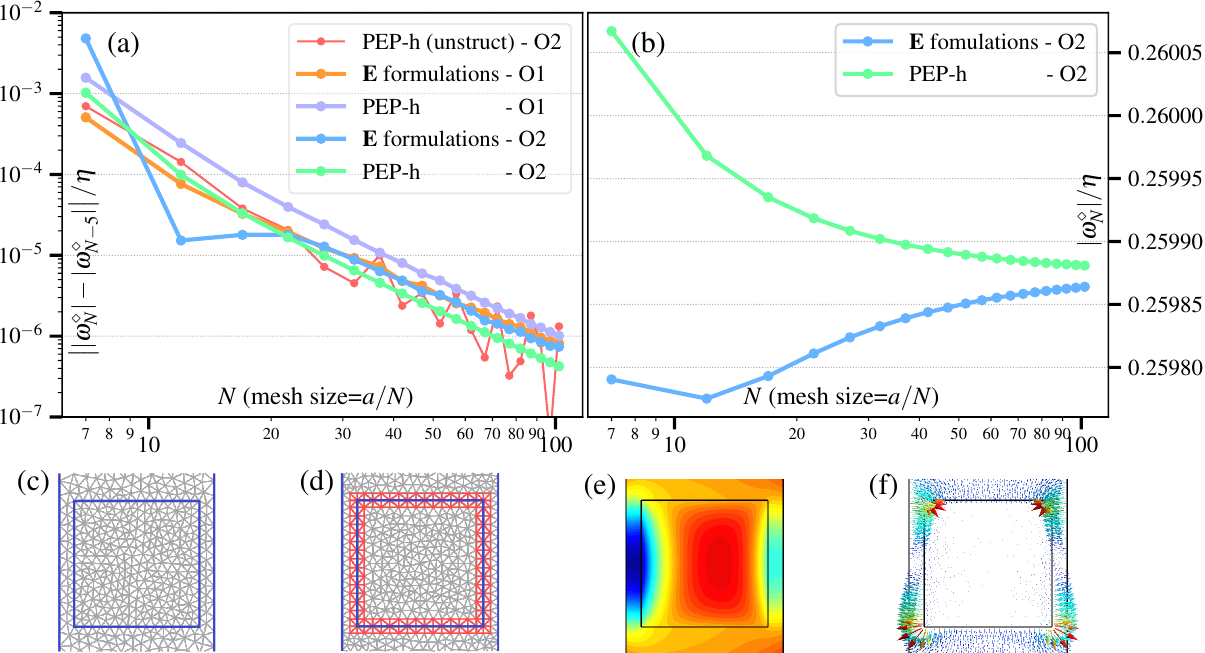}
  \caption{\crevi (a) Convergence rate of an eigenvalue of $\omfb$ as a function of
      the mesh size parametrized by $N$ for electric and magnetic formulations and two FE orders. 
      (b) $|\omfb_N|$ as a function of $N$
      (c) Unstructured mesh sample 
    . (d) Structured mesh sample 
    . (e) Real part of the scalar mode $h^{\diamond}$   corresponding to \omfb. 
      (f) Real part of the vector mode $\bE^{\diamond}$ corresponding to \omfb.}
  \label{fig:conv}
\end{figure}

Even away from the critical interval, it is legitimate to question the convergence of the eigenvalues. Let us
focus on one eigenvalue in particular, the lowest (fundamental) eigenfrequency for $\alpha=3\pi/(4a)$, denoted
$\omfb$. It corresponds to the lowest frequency band shown in Fig.~\ref{fig:eigs_cplane}(f).
Figure~\ref{fig:conv}(b) shows the value of $|\omfb|$ as a function of the mesh refinement.
For 100 mesh elements per period, 5 significant digits are found on the real part and 6 on the imaginary part.

The convergence rate of this eigenvalue with the mesh refinement is shown in
Fig.~\ref{fig:conv}(a). The numerical value of this eigenfrequency for a mesh size parametrized by $N$
is denoted $\omfb_N$ and the quantity $\big||\omfb_{N}|-|\omfb_{N-2}|\big|$ is
represented as a function of $N$, from $N=7$ to $N=100$ for
interpolation orders~1 and 2. All the formulations are represented with the following 
color code : PEP-h in purple (FE order 1) and green (FE order 2), electric field formulations in orange (FE order 1)
and blue (FE order 2). Again, it is stressed that the eigenvalues $\omfb_N$ shown in Fig.~\ref{fig:conv}(a) are identical
up to the solver tolerance irrespectively of the electric field formulation and in spite of the
different treatment of the non-linearity leading to the discrete systems (except for the PEP-E and NEP-E
cases which share the very same FE matrices).

With the classical unstructured Delaunay mesh (see Fig.~\ref{fig:conv}(c)), an erratic
behavior is obtained (see the thin red line in Fig.~\ref{fig:conv}(a)) which is consistent
with the results in \cite{chesnel2013t,carvalho2015etude}. 

The corresponding modes profiles $h^{\diamond}$ (obtained with nodal elements and the PEP-h approach) 
and $\bE^{\diamond}$ (obtained with edge elements and the NEP-E approach) are depicted in
Fig.~\ref{fig:conv}(e) and Fig.~\ref{fig:conv}(f) respectively. It is clear from this last figure
that the hot spots at the corners play an important role in the convergence, even though $\omfb$ is away from
the critical interval. It can explain the absence of the expected change of slope \cite{geuzaine1999convergence} 
in the convergence rate when increasing the polynomial order of the FE shape functions: The solution at corners is not
regular enough to be significantly improved by higher orders.

\subsection{Computation time and memory requirements}
Some computation details are given for the most time consuming simulations presented in this paper
used to produce Figs.~\ref{fig:eigs_cplane}(a) with $N=8$ and second order FE. 
These simulations ran on a machine equipped with Intel Xeon 2.7GHz processors. 
First, the RAM memory used is {\crevme linked to} both the system size and the SLEPc solver
used. The most memory and time consuming approach is the auxiliary field one. The extra volume
unknowns increase the system size by one third compared to the PEP/NEP-E approaches. 
Note that there is one single auxiliary field in this Drude case. 
The average computation times for one value of the Bloch variable are, by descending order, 
$10\mathrm{m}18\mathrm{s}$ (Aux-E, 44596 DOFs, 2.8Gb memory), 
$9 \mathrm{m}45\mathrm{s}$ (Lag-E, 34750 DOFs, 2.6Gb memory), 
$9 \mathrm{m}30\mathrm{s}$ (PEP-E, 34150 DOFs, 2.6Gb memory), 
$8 \mathrm{m}49\mathrm{s}$ (NEP-E, 34150 DOFs, 1.5Gb memory), 
$5 \mathrm{m}13\mathrm{s}$ (PEP-h, 23688 DOFs, 2.3Gb memory).
The computation times follows unsurprisingly the number of degrees of freedom and the sparsity 
of the matrices. Note that the approach using SLEPc non-linear rational NLEIGS solver 
is the fastest for this problem among vector formulations, faster than PEP-E, due to its smaller
memory footprint.

Finally, with a reasonably fine mesh ($N=6$), it is stressed that the provided model that retrieves 
the eigenvalue $\omfb$ runs within a few seconds on any laptop.



\begin{table}[]\footnotesize
  \centering
  \label{tab:advantage}
  \caption{A synthetic view of strengths and weaknesses of all the presented approaches.}
  \begin{tabular}{|l|l|l|}
    \hline
    Approach &  Advantage  &  Limitation  \\ \hline
    Aux-E    &  
        \tabmr{\textbullet~Physical linearization \\
               \textbullet~Low polynomial order (2) \\
               \textbullet~Easy to extend to several \\
              \wtextbullet~materials with more poles} &  
        \tabmr{\textbullet~System size \\
               \textbullet~Speed} \\ \hline
    PEP-E    &                      
        \tabmr{\textbullet~Smallest system size} &
        \tabmr{\textbullet~Tedious to generalize \\
               \textbullet~Polynomial order\\ 
              \wtextbullet~with several materials} \\\hline
    NEP-E    &                                      
        \tabmr{\textbullet~Smallest system size \\
               \textbullet~Memory footprint \\
               \textbullet~Shortest runtime \\
               \textbullet~Ease of implementation} &
        \tabmr{\textbullet~Stability and \\ 
              \wtextbullet~convergence with \\
              \wtextbullet~several materials?} \\\hline
    Lag-E    &
        \tabmr{\textbullet~Domain by domain \\
              \wtextbullet~formulation with extra\\
              \wtextbullet~boundary unknowns  \\
               \textbullet~Low polynomial order } &  
        \tabmr{\textbullet~Extra boundary \\
              \wtextbullet~unknowns} \\\hline
    PEP-h    &
        \tabmr{\textbullet~For comparison purposes, \\ 
              \wtextbullet~especially around \\
              \wtextbullet~the critical interval} &  
        \tabmr{\textbullet~2D scalar case only} \\
    \hline
  \end{tabular}
\end{table}

\cblack
\section{Conclusion} 

In this paper, we have presented a framework to solve non-linear eigenvalue problems suitable {\crevi for} a Finite
Element discretization. The implementation is based on the open-source FE software GetDP and
the open-source library SLEPc.

Several approaches aimed at the linearization of the eigenvalue problem arising from the consideration of
frequency-dispersion in electromagnetic structures have been introduced, implemented and discussed. The
relative permittivity was considered under the form of a rational function of the eigenvalue with arbitrary
degrees for the denominator and numerator. Five formulations were derived in the frame of a typical
multi-domain problem exhibiting several key features in electromagnetism: The mono-dimensional grating is a
quasi-periodic problem with PMLs. This is a 2D problem quite representative of 3D situations since the
physics is as rich as in 3D (exhibiting surface plasmons) and the vector case with edge elements is
tackled. We take advantage of the performance and versatility of the SLEPc library whose non-linear
eigenvalue solvers were interfaced with the flexible GetDP FE GNU software for the purpose of
this study. {\ced An open-source template model based on the ONELAB interface to Gmsh/GetDP is provided and
can be freely downloaded from \cite{nleig_code}. It exhibits the various ways to set up non-linear EVPs in
the newly introduced GetDP syntax: One rational EVP and four polynomial EVPs with various degrees are
shown.}

The first four formulations of the 2D grating problem concern the vector case and the choice of unknown
is the electric field. First, physical auxiliary fields (Aux-E) allow to linearize of the problem by
extending Maxwell's operator. The unknowns are added in the dispersive domains solely. The final
polynomial EVP is quadratic. Second, writing the Maxwell problem under its variational form brings out a
rational (NEP-E) and a polynomial (PEP-E) eigenvalue problem. An alternative consists in dealing with the
rational function under the strong form of the problem and making the use of Lagrange multipliers (Lag-E)
to deal with the non-classical boundary terms arising from this formulation. The advantage of this
approach is to keep the order of the polynomial EVP as small as possible. Finally, for comparison, the polynomial
approach is given for the scalar version of same polarization case using the magnetic field (PEP-h).

We obtain a perfect numerical agreement between all the electric field approaches in spite of the fact
that they rely on very different linearization strategies. As for the magnetic one, when {\crevme away from the
critical interval} inherent to the presence of the sign changing permittivity and sharp angles,
the agreement still holds. As for this critical interval associated with solutions of infinite energy,
they cannot be captured with a classical FE scheme. Specific PMLs could be adapted. However,
away from the critical interval, for instance for the fundamental mode of the grating, a smooth
convergence is obtained when using a specific locally structured and symmetric mesh.

{To conclude on the main features of the presented approaches, the SLEPc rational NLEIGS solver
used in the NEP-E approach gives the best results in terms of ease of implementation, speed, and memory
occupation in this test case with a simple Drude model. In spite of its much larger size than with all
other approaches, the auxiliary fields approach is very valuable for validation purposes since it relies
on a very different linearization mechanism and thus completely different sparse matrices. The approach
using Lagrange multipliers (Lag-E) deserves some attention since the polynomial order will not {\crevme blow up with}
an increased number of dispersive materials. Finally, for this problem involving the permittivity
directly given as a rational function, the NEP-E approach should be preferred over the PEP-E one.

Direct perspectives of this work consist in applying {\crevme these different} approaches to the 3D case
with more sophisticated permittivity functions. But considering several distinct materials, relying on
permittivity functions with more poles, implies the presence of more complicated frequency lines in the
complex plane leading to additional critical intervals. Inevitably, special {\crevme treatments} should be
investigated for the corners issue.

\appendix
\renewcommand{\thesection}{Appendix \Roman{section}}
\renewcommand{\thesubsection}{\roman{subsection}}
\renewcommand{\thefigure}{\arabic{figure}}
\renewcommand{\thetable}{\arabic{table}}
\crevi
\section{A standard representation of the dispersion relation}\label{ap:disp}
\cblack
{\crevi A standard representation of the dispersion relation for gratings is shown in Fig.~\ref{fig:eigs_real_imag}
using the very same numerical data as in Fig.~\ref{fig:eigs_cplane}. This figure shows
Re$\{\omega\}$ against $\alpha$, while Im$\{\omega\}$ is given in color scale.}

\begin{figure}[h]
  \centering
  \includegraphics[width=.92\textwidth,draft=\flagdraftfig]{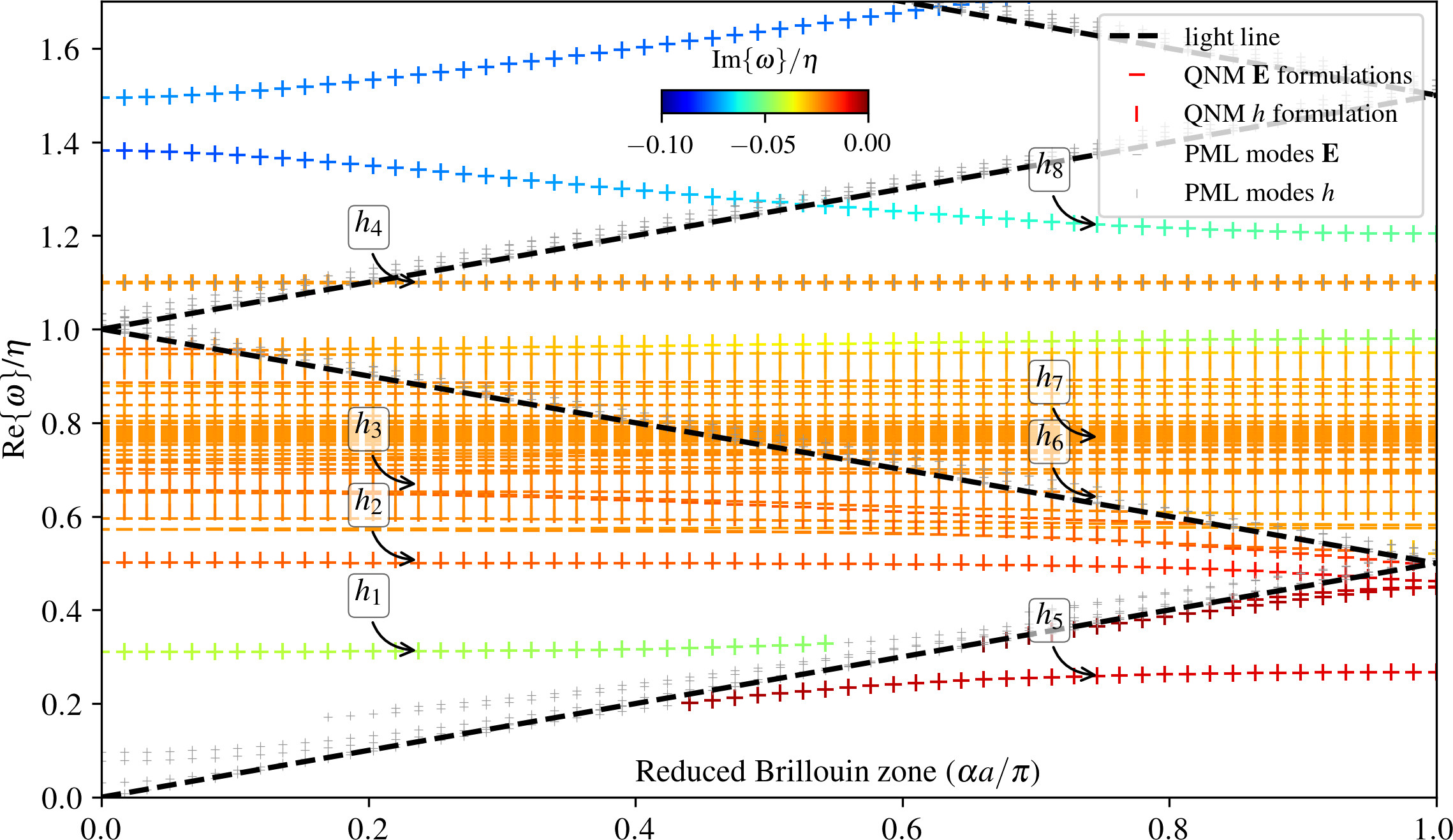}
    \caption{(a) Real (ordinate axis) and imaginary parts (jet color scale) of the normalized 
             eigenvalues ($\eta=\frac{2\pi c}{a}$) as a function of the Bloch variable $\alpha$ for the five methods. 
             Electric (resp. magnetic) field formulations are represented by hyphens \enquote{{\tt\_}} (resp. vertical hyphens \enquote{{\tt|}}). 
             The folded light line is depicted as a black dashed line.}
    \label{fig:eigs_real_imag}
\end{figure} 

{\crevi The modes of the continuum (or free space modes or PML modes) corresponding to radiation losses are shown in
grey symbols for both formulations. Two criteria are used to classify these modes as PML modes. 
The other modes are considered as QNMs of the grating. The first criterion relies on the independence of the QNM 
towards the PML parametrization. The dispersion relation has been computed twice with two different values of the complex coordinate stretch
parameter $s_y$ (1+i and 1+2i) defined in Eq.~(\ref{eq:epspml}). The eigenvalues whose both real and imaginary parts change
by less than $1\%$ between the two computations are kept and fed to the next criterion. Indeed the first stability criterion
is sufficient for an isolated scatterer surrounded by a PML: A single branch of continuous spectrum is rotated around the
origin by an angle of Arg$\{s_y\}$/2. However, in periodic cases, several PML branches are obtained in the frequency range of
interest, which corresponds to the fact that the structure interacts with the continuum through its infinite set of
diffraction orders \cite{vial_quasimodal_2014}. These branches rotate by an angle of Arg$\{s_y\}$/2 around the points sitting
at $n\pi/a$ on the real line, where $n$ is an integer. As a consequence, all the PML modes close to these points on the
real line are not discarded by the first criterion above. The second criterion relies on the fact that eigenvectors
corresponding to PML modes are mostly located into the PMLs as shown in Fig.~\ref{fig:eigs_cplane}(g). The second criterion classifies as PML mode an eigenvector $h_n$ satisfying 
$
\int_{\Omega_1^d\cup\Omega^\toslash_2}|h_n|\,\mathrm{d}\Omega /
\int_{\Omega_1^\toslash\cup\Omega^\toslash_3}|h_n|\,\mathrm{d}\Omega >0.5$. Note that the
threshold values of $1\%$ for the first criterion and 0.5 for the second criterion depend on the mesh refinement and PML
thicknesses respectively. Modes which do not fall into the two categories defined above are considered as QNMs and represented
by colored vertical hyphens (PEP-h) and horizontal hyphens (electric field cases) in Fig.~\ref{fig:eigs_real_imag}.

Just below the first branch of the folded light line represented by the dashed black line, the shape of the band corresponding
to the lowest eigenfrequency supported by the grating is characteristic of the fundamental mode of this type of structure
\cite{lalanne2005surface,lalanne2006optical,schider2003plasmon,lalanne2019quasinormal}. The real part of an eigenfield of 
this particular band ($h_5$) is shown in Fig.~\ref{fig:eigs_cplane}(g). Other higher bands appear below the first branch of the folded 
light line for higher values of the Bloch wavevector $\alpha$. After the first folding of the light line at 
Re$\{\omega\}/\eta\approx0.45$, classical bands are retrieved but some discrepancy appears between the electric formulations and 
the magnetic one as detailed in Sec.~\ref{sec:dispcplane}. 
The accumulation of flat bands in the range $0.55<\mathrm{Re}\{\omega\}/\eta<1.1$ corresponds to plasmons and corner modes. Finally, 
the dispersion relation retrieves a more conventional behavior, with very leaky higher frequency modes 
such as mode $h_8$ in Fig.~\ref{fig:eigs_cplane}(i).}
\creviii
\section{Structured meshes and rounding} \label{ap:round}
\cblack
\begin{figure}[h!] \centering
  \includegraphics[width=.92\textwidth,draft=\flagdraftfig]{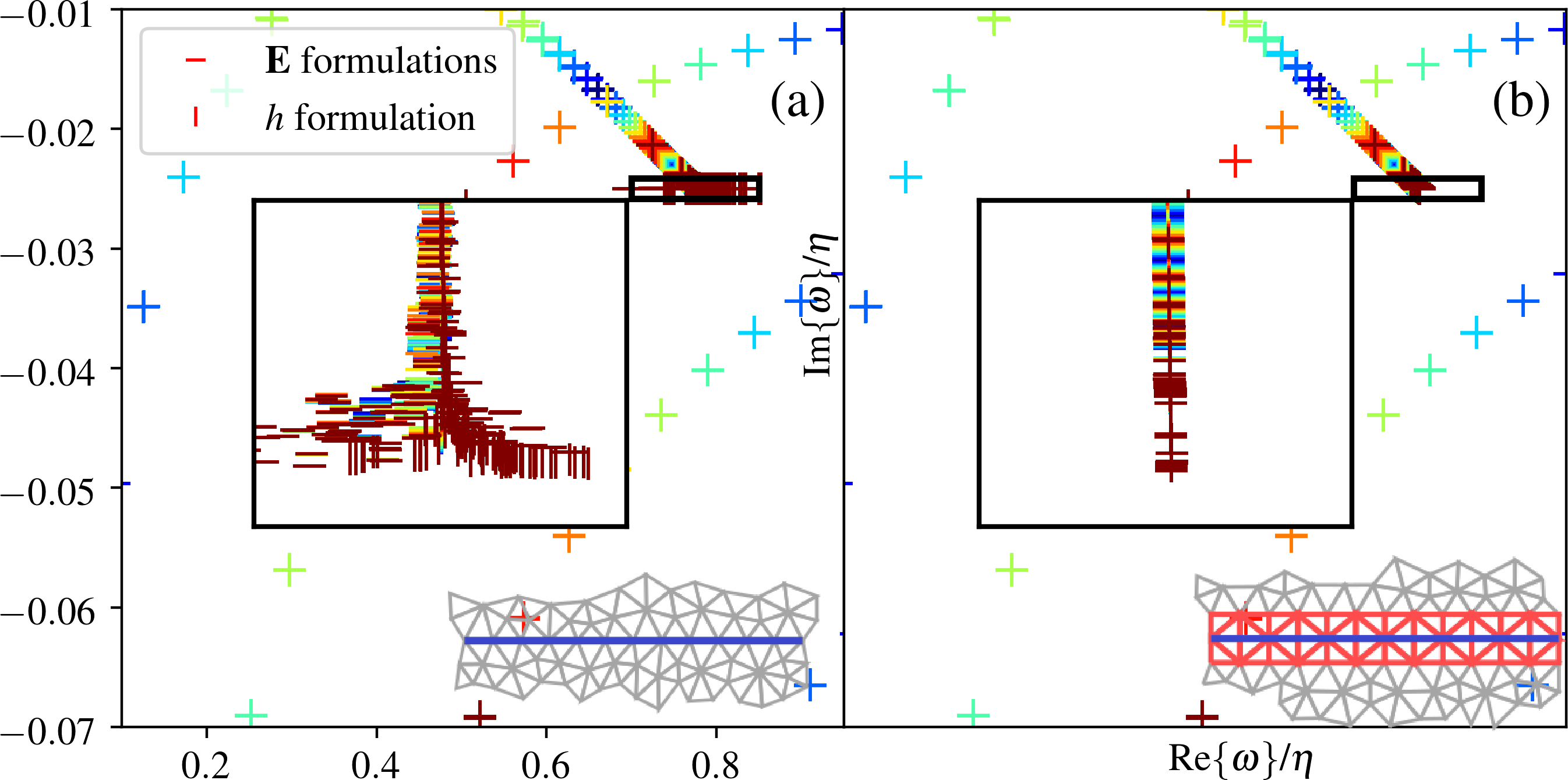}
  \caption{\crevme Spectrum of the slab in the complex plane calculated in 2D with a classical unstructured mesh 
  (a) and with a structured mesh (b).}
  \label{fig:flat}
\end{figure}  
{\crevme To illustrate the importance of imposing a symmetric mesh around a flat interface, the spectrum of a slab is depicted in \Fig{fig:flat}. 
The slab is a 1D problem that can be treated here in 2D since it is trivially periodic (b=a). An accumulation point is expected
at $\omega$ such that $\varepsilon_{r,1}(\omega)=-1$ (framed by a black rectangle in Figs.~\ref{fig:flat}(a-b)).
As can be noticed in the two insets, it is striking that the unstructured mesh leads to a poor description of
the plasmonic accumulation point where the structured symmetric mesh preserves stability.}
\begin{figure}[h!] \centering
  \includegraphics[width=.92\textwidth,draft=\flagdraftfig]{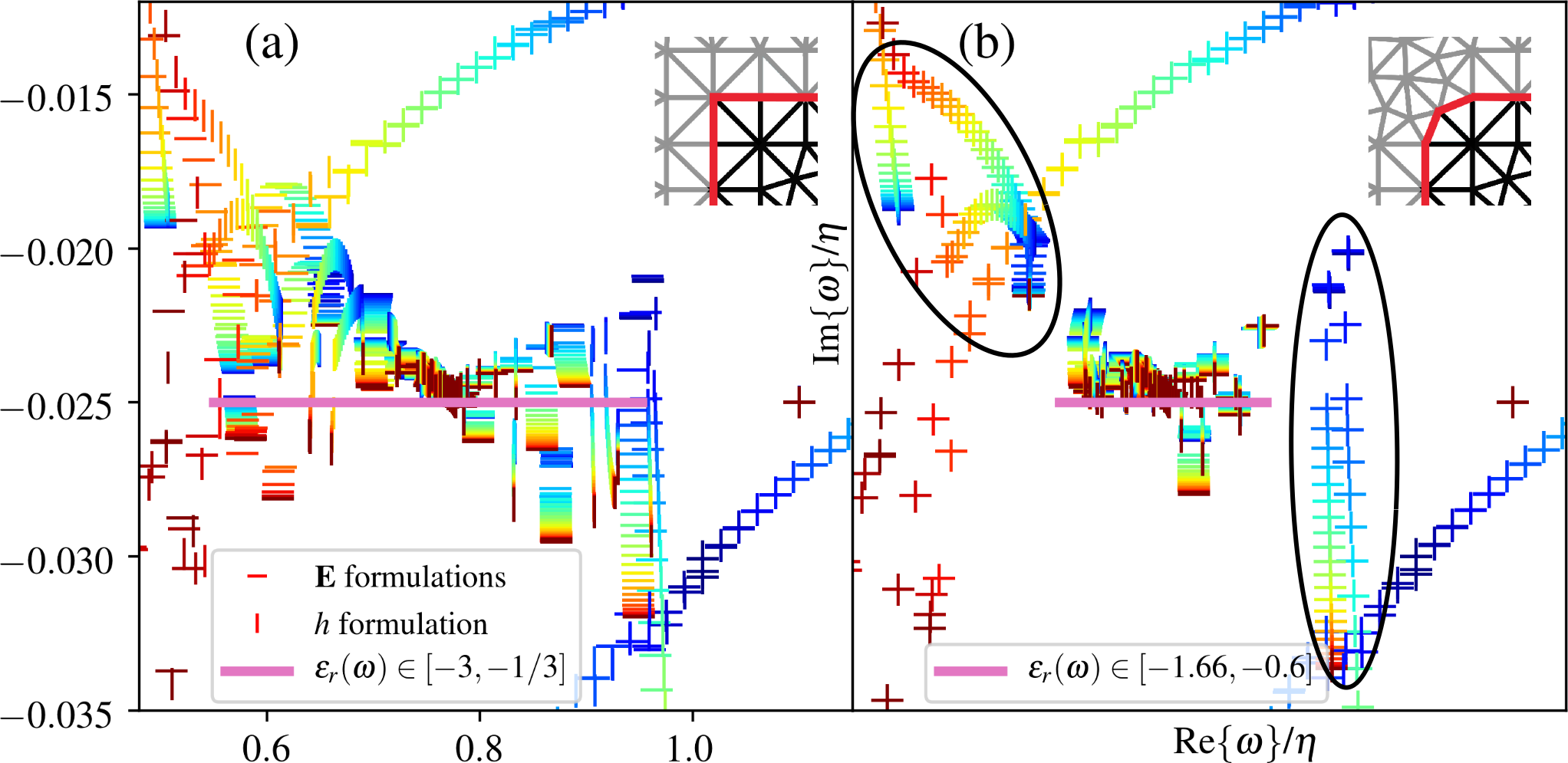}
  \caption{\creviii (a) Zoom in the black rectangle of Fig.~\ref{fig:eigs_cplane}(a). (b) Same result for a 
  slight rounding of radius  $a/25$ of all corners, \emph{i.e.} a critical angle of 
  $3\pi/4$ on three mesh nodes (see inset of the right pannel). }
  \label{fig:rounding}
\end{figure}
{\creviii The agreement between electric and magnetic formulations can be improved by slightly rounding the corners. 
The left panel of Fig.~\ref{fig:rounding} corresponds to a zoom in the black frame of Fig.~\ref{fig:eigs_cplane}.
The discrepancy is clearly visible. The right panel shows the spectrum of the slightly rounded square 
(the radius of the rounding is $a/25$, cf. mesh sample in the right inset). From the discrete point of view, 
four consecutive mesh edges of the mesh along the rounded corner form two by two angles of $3\pi/4$, which corresponds to
a critical interval for $\eps{r}{1}(\om)$ of $[-1.66,-0.6]$ (pink line in the right panel). The agreement between 
the electric and magnetic formulations is now striking away from the reduced critical region (cf. hyphens \enquote{{\tt\_}} 
and \enquote{{\tt|}} now forming \enquote{{\tt+}} in the oval black frames).}

\section{Implementation in GetDP}\label{ap:getdp}
{\ced 
  The GetDP software is an open source FE solver (\spath{http://getdp.info}). It
  handles geometries and meshes generated using the open source mesh generator Gmsh
  (\spath{http://gmsh.info}). The source codes of both softwares are available at
  \spath{https://gitlab.onelab.info}. 
  
  A template model to allowing to retrieve the results of this paper is also available \cite{nleig_code}. It relies on ONELAB, a
  lightweight interface between Gmsh and GetDP. To run the example, one can simply (i) download the
  precompiled binaries of Gmsh and GetDP available for all platforms as a standalone ONELAB bundle from
  \spath{http://onelab.info}, (ii) download the template model and (iii) open the
  \stexttt{NonLinearEVP.pro} file with Gmsh.
  
  This work has involved changes to GetDP in both the source code and the parser in order to call the
  relevant SLEPc solvers in a general manner. These changes now allow to solve a large class of
  non-linear (polynomial and rational) eigenvalue problems suitable for a FE discretization.
  Indeed, the software readily handles various FE basis functions relevant in
  electromagnetism, acoustics, elasticity\dots The example in electromagnetism in this paper has
  voluntarily been chosen relatively simple for the sake of clarity. As shown in Sec.~\ref{sec:corners}
  both the computation and the underlying physics of dispersive gratings modes are rather intricate.

  In practice, a problem definition written in \stexttt{.pro} input files is usually split
  between the objects defining data particular to a given problem, such as geometry,
  physical characteristics and boundary conditions (i.e., the \stexttt{Group},
  \stexttt{Function} and \stexttt{Constraint} objects), and those defining a resolution
  method, such as unknowns, equations and related objects (i.e., the \stexttt{Jacobian},
  \stexttt{Integration}, \stexttt{FunctionSpace}, \stexttt{Formulation}, \stexttt{Resolution}
  and \stexttt{PostProcessing} objects). The processing cycle ends with the presentation of
  the results, using the \stexttt{PostOperation} object.
  
  The major changes appear at \stexttt{Formulation} and \stexttt{Resolution} stages. A new \stexttt{Eig} operator was
  introduced in the parser. It can be invoked to set up a polynomial EVP when combined with the keyword
  \stexttt{Order}, or a rational EVP when combined with the keyword \stexttt{Rational}. The \stexttt{Order} or
  \stexttt{Rational} keywords allow to define the dependence of the problem with the eigenvalue $\lambda:=i\omega$. Depending
  on whether \stexttt{Order} or \stexttt{Rational} is set, GetDP internally calls the static functions
  \stexttt{\_polynomialEVP} or \stexttt{\_nonlinearEVP} where the interface to SLEPc is written in practice. These functions can be
  found in the source code of GetDP in the C++ file \spath{Kernel/EigenSolve_SLEPC.cpp} for further details.
  
  Note that in all GetDP eigenvalue solvers the eigenvalue has been chosen to be $i\omega$, consistently with
  the convention in this paper. In the following GetDP listings, the dots (\textbf{...}) represent a 
  deliberate omission of some instructions that are unnecessary to the comprehension of the implemented syntax.
  The reader is invited to refer to the template example to see these GetDP code snippets in their global context.
  
  \subsection{Polynomial eigenvalue problems}
  \begin{table}[h]
    \centering \small
    \caption{Correspondence between mathematical and GetDP objects.}
    \label{tab:corresp}
    \begin{tabular}{|l|l|l|}
      \hline
      GetDP object & Mathematical object & Description \\
      \hline 
      \spath{cel}       &  $c$                          & light celerity \\
      \spath{I[]}       &  $i$                          & $i^2=-1$       \\
      \spath{mur[]}     &  $\tensmur(\br)$              & Tensor field \\
      \spath{epsr_nod[]}&  $\tensepsr^{\toslash}(\br)$  & Tensor field \\

      \spath{eps_oo_1}  &  $\epsinf$ (cf. Eq.~(\ref{eq:epsdrude1})) & Flat contribution \\
      \spath{om_d_1}    &  $\omd$ (cf. Eq.~(\ref{eq:epsdrude1}))    & Plasma frequency     \\
      \spath{gam_1}     &  $\gamd$ (cf. Eq.~(\ref{eq:epsdrude1}))   & Damping frequency     \\
      
      \spath{Om}        &  $\Omega$           & Computational domain   \\
      \spath{Om_1}      &  $\Omega_1^d$       & Dispersive domain      \\
      \spath{Om_2}      &  $\Omega^\toslash$  & Non-dispersive domains \\
      \hline
      \lstinline|Galerkin{ [ Dof{Curl u}, | & & Contribution to the \\
      \lstinline|  {Curl u}]; In Om ; ... }| & $+\int_{\Omega} \curl\,\bE\cdot\overline{\curl\,\bW}\,\mathrm{d}\Omega$ & variational formulation \\
      \hline
      \lstinline|Galerkin{ Eig[ Dof{u}, {u}];| & & Contribution to the \\
      \lstinline|  Order 3 ; In Om_1  ; ... }| & $+\lambda^3\int_{\Omega_d^1} \bE\cdot\overline{\bW}\,\mathrm{d}\Omega$ & variational formulation in $\lambda^3$ \\
      \hline
    \end{tabular}
  \end{table}   
GetDP now solves polynomial eigenvalue problems. Its syntax is shown in the listing~\ref{lst:pep}. This 
GetDP formulation corresponds to the PEP-E formulation mathematically described in Eq.~(\ref{eq:PEP-E}).
For clarity, the correspondence between the relevant mathematical objects and GetDP objects are detailed in 
Table~\ref{tab:corresp}.

\begin{lstlisting}[caption={Syntax for the formulation of the polynomial eigenvalue problem. 
  The dots (\textbf{...}) represent a  deliberate ellipsis to the code.},label={lst:pep}]
{ Name pep; Type FemEquation;
  Quantity {
    { Name u ; Type Local; NameOfSpace Eedge;}
  }
  Equation {
    Galerkin{   [-cel^2/mur[]*gam_1*Dof{Curl u},{Curl u}];          In Om  ; ...}
    Galerkin{Eig[ cel^2/mur[]      *Dof{Curl u},{Curl u}]; Order 1; In Om  ; ...}
    Galerkin{Eig[ om_d_1^2         *Dof{u}     ,{u}     ]; Order 1; In Om_1; ...}
    Galerkin{Eig[-eps_oo_1*gam_1   *Dof{u}     ,{u}     ]; Order 2; In Om_1; ...}
    Galerkin{Eig[ eps_oo_1         *Dof{u}     ,{u}     ]; Order 3; In Om_1; ...}
    Galerkin{Eig[-epsr_nod[]*gam_1 *Dof{u}     ,{u}     ]; Order 2; In Om_2; ...}
    Galerkin{Eig[ epsr_nod[]       *Dof{u}     ,{u}     ]; Order 3; In Om_2; ...}
  }
}
\end{lstlisting}
Note that the PEP-h formulation involves a 4$^{th}$ order polynomial eigenvalue problem, and the Aux-E formulation
involves a quadratic one.

  \subsection{Rational non-linear eigenvalue problems}
  GetDP now solves rational eigenvalue problems. Its syntax is shown in Listing~\ref{lst:nep}.  This 
  GetDP formulation corresponds to the NEP-E formulation mathematically described in Eq.~(\ref{eq:NEP-E}).
\begin{lstlisting}[caption={Syntax for the formulation of the rational eigenvalue problem.
  The dots (\textbf{...}) represent a  deliberate ellipsis to the code.},label={lst:nep}]
{ Name form_nep; Type FemEquation;
  Quantity {
    { Name u ; Type Local; NameOfSpace Eedge;}
  }
  Equation {
      Galerkin{Eig[ cel^2/mur[]*Dof{Curl u}, {Curl u} ]; Rational 1; In Om  ; ... }
      Galerkin{Eig[-epsr_nod[] *Dof{u}     , {u}      ]; Rational 2; In Om_1; ... }
      Galerkin{Eig[-epsr_nod[] *Dof{u}     , {u}      ]; Rational 3; In Om_2; ... }
  }
}
\end{lstlisting}
Then, at the \stexttt{Resolution} step, each rational function expected as a factor of each
\stexttt{Galerkin} term is specified. The $6^{th}$ (respectively $7^{th}$) argument 
of the \stexttt{EigenSolve} function is a list of polynomial numerators (resp. denominators), 
each polynomial numerator  (resp. denominator) being itself given as a list of GetDP floats. 
The position of each numerator (resp. denominator) in the list of numerators (resp. denominators) 
corresponds to the tag following the \stexttt{Rational} keyword. A polynomial numerator 
(resp. denominator), is represented by a list of (real) floats by decreasing power of $\lambda$. 
For instance, the list \spath{{-eps_oo_1,gam_1*eps_oo_1, -om_d_1^2,0}} in Listing~\ref{lst:nepres} 
represents the polynomial $-\epsinf\lambda^3+\gamd\epsinf \lambda^2-\omd^2\lambda$, numerator of 
$\lambda^2\eps{r}{1}(\lambda)$. Likewise, the list \spath{{1,-gam_1}} in Listing~\ref{lst:nepres} 
represents the polynomial $\lambda-\gamd$, denominator of $\lambda^2\eps{r}{1}(\lambda)$. 
Note that the degrees of the numerators and denominators can be arbitrarily large.
\begin{lstlisting}[caption={Syntax for the resolution of the rational eigenvalue problem},label={lst:nepres}]
{ Name res_nep;
  System{{ Name M; NameOfFormulation form_nep; Type ComplexValue;}}
  Operation{
    GenerateSeparate[M1];
    EigenSolve[M,neig,target_real,target_imag,EigFilter[],
      {{1}, {-eps_oo_1,gam_1*eps_oo_1, -om_d_1^2,0}, {-1,0,0}} ,
      {{1}, {1,-gam_1},                              {1}     } ];
  }
}
\end{lstlisting}

  \subsection{Specifying the eigensolver}
  The general SLEPc options for solving of non-linear problems are preset in the source code of
  GetDP (see \stexttt{Kernel/EigenSolve\_SLEPC.cpp}). Additional or alternative SLEPc options can be 
  passed as command line argument when calling GetDP.
  There are particularly relevant options that can be passed to SLEPc:
\begin{itemize}
   \item \textbf{Target}: SLEPc eigensolvers will return \spath{nev} eigenvalues closest to a given
   target value. The \spath{nev} parameter can be specified by the user (1 by default), as well as the
   target value, that represents a point in the complex plane around which the eigenvalues of interest
   are located. The values can be provided via the ONELAB dialog boxes of the provided open-source model,
   or alternatively with the command line
   arguments \spath{-pep_nev} (or \spath{-nep_nev}), and \spath{-pep_target} (or 
   \spath{-nep_target}).
   \item \textbf{Regions}: The eigenvalues are returned sorted according to their distance to the target.
   However, only eigenvalues lying inside the region of interest are returned (in other words,
   eigenvalues outside the region of interest are discarded). The region of interest (which is a
   rectangle by default) can be specified by the user
   via the ONELAB dialog boxes of the provided open-source model, or alternatively with the
   command line argument \spath{-rg_interval_endpoints} (or any other options related to region
   specification, see SLEPc documentation \cite{slepc-users-manual} for details).
 \end{itemize}
    
  \subsection{Generalization}
  With the change made to GetDP, one can tackle much more general problems.
  For instance, if the geometry has N dispersive materials with distinct material dispersion, one would just need to extend 
  the recipe above, as schematized in the GetDP Listing~\ref{lst:nepgeneral}. Note that 
  a numerator or denominator can be provided as a GetDP list directly, defined in the \texttt{Function} object.
\begin{lstlisting}[caption={Syntax for a general problem with several dispersive materials},label={lst:nepgeneral}]
{ Name form_nep; Type FemEquation;
  Quantity {
    { Name u ; Type Local; NameOfSpace Eedge;}
  }
  Equation {
    Galerkin { Eig[   1/mur[] *    Dof{Curl u}, {Curl u} ]; Rational 1; In Om  ; ... }
    Galerkin { Eig[ -epsr_nod[]/cel^2 * Dof{u}, {u}      ]; Rational 2; In Om_1; ... }
    Galerkin { Eig[ -epsr_nod[]/cel^2 * Dof{u}, {u}      ]; Rational 3; In Om_2; ... }
    Galerkin { Eig[ -epsr_nod[]/cel^2 * Dof{u}, {u}      ]; Rational 4; In Om_3; ... }
    Galerkin { Eig[ -epsr_nod[]/cel^2 * Dof{u}, {u}      ]; Rational 5; In Om_4; ... }
    ...
  }
}
...
{ Name res_nep;
  System{{ Name M; NameOfFormulation form_nep; Type ComplexValue;}}
  Operation{
    GenerateSeparate[M1];
    EigenSolve[M,neig,target_real,target_imag,EigFilter[],
                {num_1(), num_2(), num_3(), num_4(), num_5(), ... }
                {den_1(), den_2(), den_3(), den_4(), den_5(), ... }];
  }
}
\end{lstlisting}
}

\section*{Acknowledgements}
The work was partly supported by the French National Agency for Research (ANR) under the
project “Resonance” (ANR-16-CE24-0013). The authors acknowledge the members of the project
“Resonance” for fruitful discussions. C.~Campos and J.~E.~Roman were supported by the Spanish
Agencia Estatal de Investigaci{\'o}n (AEI) under project SLEPc-HS (TIN2016-75985-P), which
includes European Commission ERDF funds. C.~Geuzaine was supported by ARC grant for Concerted Research 
Actions (ARC WAVES 15/19-03), financed by the Wallonia-Brussels Federation of Belgium.

The authors thank Christian Engstr{\"o}m from Ume\.a Universitet for helpful comments.
Maxence Cassier from Institut Fresnel is acknowledged. Finally, the authors address special thanks to 
Anne-Sophie Bonnet Ben-Dhia and Camille Carvalho from INRIA (POEMS) for their motivating remarks and insights. 
\bibliographystyle{ieeetr}
\bibliography{biblio}

\end{document}